\documentclass[twocolumn,trackchanges]{aastex631}

\usepackage{graphicx}
\usepackage{xcolor}
\usepackage{natbib}
\usepackage{amsmath}
\usepackage[subfigure]{tocloft}
\usepackage{subfigure}
\usepackage{booktabs}
\usepackage{CJK}

\newcounter{RomanNumber}

\received{xxx}
\revised{xxx}
\accepted{xxx}

\shorttitle{Tidal Stripping of a White Dwarf by an Intermediate-Mass Black Hole}
\shortauthors{Chen, Shen, \& Liu}

\begin{document}
\begin{CJK*}{UTF8}{gbsn}

\title{Tidal Stripping of a White Dwarf by an Intermediate-Mass Black Hole}

\author[0000-0002-3525-791X]{Jin-Hong Chen (陈劲鸿)}
\affiliation{School of Physics and Astronomy, Sun Yat-sen University, Zhuhai 519082, China}
\author[0000-0001-5012-2362]{Rong-Feng Shen (申荣锋)}
\affiliation{School of Physics and Astronomy, Sun Yat-sen University, Zhuhai 519082, China}
\affiliation{CSST Science Center for the Guangdong-Hong Kong-Macau Greater Bay Area, Sun Yat-sen University, Zhuhai 519082, China}
\author[0000-0002-9442-137X]{Shang-Fei Liu (刘尚飞)}
\affiliation{School of Physics and Astronomy, Sun Yat-sen University, Zhuhai 519082, China}
\affiliation{CSST Science Center for the Guangdong-Hong Kong-Macau Greater Bay Area, Sun Yat-sen University, Zhuhai 519082, China}
\email{shenrf3@mail.sysu.edu.cn,
liushangfei@mail.sysu.edu.cn}


\begin{abstract}
During the inspiralling of a white dwarf (WD) into an intermediate-mass black hole ($\sim 10^{2-5}\ M_{\odot}$), both gravitational waves (GWs) and electromagnetic (EM) radiation are emitted. Once the eccentric orbit's pericenter radius approaches the tidal radius, the WD would be tidally stripped upon each pericenter passage. The accretion of these stripped mass would produce EM radiation. It is suspected that the recently discovered new types of transients, namely the quasi-periodic eruptions and the fast ultraluminous X-ray bursts, might originate from such systems. Modeling these flares requires a prediction of the amount of stripped mass from the WD and the details of the mass supply to the accretion disk. We run hydrodynamical simulations to study the orbital parameter dependence of the stripped mass. We find that our results match the analytical estimate that the stripped mass is proportional to $z^{5/2}$, where $z$ is the excess depth by which the WD overfills its instantaneous Roche lobe at the pericenter. The corresponding fallback rate of the stripped mass is calculated, which may be useful in interpreting the individual flaring light curve in candidate EM sources. We further calculate the long-term mass-loss evolution of a WD during its inspiral and the detectability of the GW and EM signals. The EM signal from the mass-loss stage can be easily detected: the limiting distance is $\sim 320(M_{\rm h}/10^4\ M_{\odot})^{1/2}$ Mpc for Einstein Probe. The GW signal, for the space-borne detectors such as Laser Interferometer Space Antenna or TianQin, can be detected only  within the Local Supercluster ($\sim 33$ Mpc).
\end{abstract}

\keywords{accretion, accretion disks - black hole physics - galaxies: nuclei }

\section{Introduction}
\label{sec:introduction}
Intermediate-mass black holes (IMBHs) are the link between stellar-mass ($\sim 10\ M_{\odot}$) and supermassive black holes (SMBHs, $\sim 10^{6 - 9}\ M_{\odot}$), and the possible seeds for the SMBHs growth in the early Universe. Proving their existence and the determination of their spatial and mass distribution are thus crucial for understanding the formation of SMBHs and the consequent galaxy formation \citep{Mezcua_IMBH_2017}. 

The center of star clusters may harbour IMBHs \citep{Miller_Production_2002,Portegies_The_2002}. There are few but not solid detections of IMBHs in globular clusters \citep{Gerssen_Hubble_2002,Gebhardt_A_2002}. Another promising location for finding an IMBH is the nuclei of dwarf galaxies \citep{Volonteri_The_2003,Moran_Black_2014,Baldassare_Xray_2017}, from which dozens of IMBHs candidates have been discovered \citep{Greene_IMBH_2020}.

In a globular cluster or a dwarf galaxy, the orbits of stars surrounding the IMBH are constantly subject to gravitational perturbations by other stars, which might cause a close encounter between a white dwarf (WD) and the IMBH. The less bound WD could suffer from a single-passage disruption \citep{Rees_Tidal_1988}. On the contrary, the tightly bound WD could undergo a multiple-passage mass transfer \citep{MacLeod_Illuminating_2014,MacLeod_The_2016}. The latter case could also form through binary splitting \citep{Hills_Hyper_1988}. In this paper, we focus on the tightly bound ones with eccentricity $e \sim 0.7 - 0.9$.

The periodic electromagnetic (EM) and gravitational wave (GW) signals from these systems can be detected by X-ray telescopes and next-generation gravitational wave detectors, respectively \citep{Zalamea_White_2010,Shen_Fast_2019}. Therefore, these systems offer a promising window for finding IMBHs and studying the accretion physics and testing the general relativity.

The EM radiation comes from the periodic accretion onto an IMBH, after the WD is tidally stripped near the pericenter. Recent discoveries of several quasi-periodic eruptions (QPEs) and fast ultraluminous X-ray bursts (ULXBs) have been suspected that they originate from such systems \citep{Shen_Fast_2019,King_GSN069_2020,King_QPE_2022,Wang_A_2022,Zhao_QPE_2021,Lu_QPE_2022}. The QPEs have almost symmetric profile in each outburst with duration $\sim 10^{3-4}$ s and peak luminosity $\sim 10^{41-42}\ {\rm erg\ s^{-1}}$, and have quasi-periods of $\sim 10^{4-5}$ s, \citep{Miniutti_Nine_2019,Giustini_Xray_2020,Arcodia_QPE_2021,Chakraborty_Possible_2021}. Those fast ULXBs are bright in the X-ray with fast rise ($\sim 10^{1-2}$ s) and decay ($\sim 10^{2-4}$ s), and with recurrence time $\sim$ days and peak luminosity $\sim 10^{40-42}\ {\rm erg\ s^{-1}}$ \citep{Sivakoff_Luminous_2005,Jonker_Discovery_2013,Irwin_Ultraluminous_2016}. These QPEs and fast ULXBs are located either in the galactic nuclei or in the star clusters, respectively.

Furthermore, such WD inspiral system would be the ideal targets for next-generation GW detectors, e.g., Laser Interferometer Space Antenna (LISA) \citep{Amaro_LISA_2017} and TianQin \citep{Luo_TianQin_2016}. \cite{Sesana_Observing_2008} suggested that these systems are detectable by GW within a distance $\sim 200$ Mpc. However, they only considered the nearly circular orbits with eccentricity $e \sim 0.1-0.3$. For the highly eccentric orbit ($e \gtrsim 0.9$), \cite{Chen_MHz_2022} find that their GW signals are too weak to be identified by LISA if the WD mass is small ($\sim 0.1-0.3\ M_{\odot}$), and instead they contribute to the GW background. For the WD inspiral systems with larger WD masses ($\gtrsim 0.6\ M_{\odot}$), we expect that their GW signals could be detected.

The observational EM properties of these systems, i.e., peak luminosity, rise time, and decay time, all depend on the specific orbital parameters, i.e., the pericenter radius, eccentricity, and also the WD mass and BH mass. \cite{Zalamea_White_2010} analytically calculated the stripped mass during each pericenter passage (their Eq. 7) by assuming an tidally unperturbed WD, and further calculated the evolution of the mass loss.

Stellar tidal disruption event (TDE), unlimited to WDs as the disrupted object, has been studied in many works by using the hydrodynamic simulation. \cite{Guillochon_Hydrodynamical_2013} explored the orbital dependence of the mass fallback rate in the single-passage disruptions. The disrupted star they considered is a solar-type star with a single polytrope ($\gamma = 5/3$ or $4/3$). Eccentric full TDEs have also been explored \citep{Hayasaki_Finite_2013,Bonnerot_Disc_2016,Cufari_Eccentric_2022}. In these single-passage TDEs they considered, the stars only experience one tidal encounter, after that the survived stars fly away, or the stars are fully disrupted. In contrast to the single-passage TDEs, in an eccentric tidal stripping, the star will undergo many pericenter passages. Eccentric Tidal stripping of a giant star and a solar-type star are studied in \cite{MacLeod_Spoon_2013} and \cite{Liu_Tidal_2023}, respectively. For the tidal stripping of a WD, \cite{Cheng_Relativistic_2013} explored the relativistic effects in the tidal interaction. \cite{Rosswog_Tidal_2009} studied the strong encounter between the WD and IMBH, which causes the explosive nuclear burning.

In this paper, we use hydrodynamic simulations to study the orbital dependence of the stripped mass from a WD by an IMBH and the corresponding mass fallback rate. The long-term evolution of the inspiral system, up to the final disruption of the WD, is also considered by analytical method. Furthermore, we briefly study the detectability of EM and GW signals from these systems.

In Section \ref{sec:analy_WD_tidal}, we analytically calculate the tidal stripping of the WD and the mass fallback rate. In section \ref{sec:simulation}, we present our hydrodynamic simulation results and compare them with the analytical calculations. Using these results, we calculate the mass-loss evolution of the WD in Section \ref{sec:evolution}. In Section \ref{sec:GW}, we investigate the detectability of the EM and GW signals from WD-IMBH inspiral during the mass-loss stage. We summarize and discuss the results in Section \ref{sec:conclusion}. In the Appendix \ref{subsec:LE}, we present the structure of the WD we used in this paper. Throughout the paper, we assume a standard $\Lambda$CDM cosmology with parameters $H_0 = 70\ {\rm km\ s^{-1}\ Mpc^{-1}}$, $\Omega_{\Lambda} = 0.7$, and $\Omega_{\rm M} = 0.3$.


\section{Analytical calculation of tidal stripping}
\label{sec:analy_WD_tidal}
\subsection{Tidally stripped mass}
\label{subsec:analy_strip_mass}
The stripped mass of a WD by an IMBH was calculated in \cite{Zalamea_White_2010} (see their Eq. 7). Here we re-do the calculation by using a  mass-radius relation of WD \citep{Paczynski_Models_1983}
\begin{equation}
R_* = 9 \times 10^8 \left[1-\left(\frac{M_*}{M_{\rm ch}}\right)^{4/3}\right]^{1/2} \left(\frac{M_*}{M_{\odot}}\right)^{-1/3}\ {\rm cm},
\label{eq:m_r}
\end{equation}
that is different from the simplified one in \cite{Zalamea_White_2010} (their Eq. 5). Here $M_{\rm ch} \simeq 1.44\ M_{\odot}$, $M_*$ and $R_*$ are the Chandrasekhar mass, WD's mass and radius, respectively. The average density of WD can be written as 
\begin{equation}
\overline \rho_* \simeq 6.6 \times 10^5 \left[1-\left(\frac{M_*}{M_{\rm ch}}\right)^{4/3}\right]^{-3/2} \left(\frac{M_*}{M_{\odot}}\right)^{2} \ {\rm g\ cm^{-3}}
\label{eq:rho_WD}.
\end{equation}

During the inspiralling, the WD loses angular momentum $j = \sqrt{G M_{\rm h} a (1-e^2)}$ by GW radiation, where $M_{\rm h}$, $a$ and $e$ are the BH mass, the semi-major axis and the orbital eccentricity, respectively. When its pericenter radius $R_{\rm p} = a(1-e)$ approaches the tidal radius $R_{\rm T} = R_*(M_{\rm h}/M_*)^{1/3}$, the WD begins to lose mass momentarily at the pericenter. Defining an impact factor as $\beta \equiv R_{\rm T} / R_{\rm p}$, we can write the pericenter radius in unit of the BH's Schwarzschild radius $R_{\rm S} = 2 G M_{\rm h} /c^2$ as
\begin{equation}
R_{\rm p} \simeq 12\ \beta^{-1} M_{\rm h,4}^{-2/3} \overline \rho_{*,5}^{-1/3} R_{\rm S}.
\label{eq:R_p}
\end{equation}
Here and after we use the conventional notation $X_n= X/10^n$, e.g., $M_{\rm h,4}= M_{\rm h}/10^4\,M_\odot$.

The orbital period of the WD is
\begin{equation}
\begin{split}
P &= 2\pi \sqrt{\frac{a^3}{G M_{\rm h}}} \\
&= 1.2 \times 10^3\ \left( \frac{\beta}{0.5} \right)^{-3/2} \left( \frac{1-e}{0.1} \right)^{-3/2} \overline{\rho}_{*,5}^{-1/2}\ {\rm s}.
\end{split}
\label{eq:t_WD}
\end{equation}

At the pericenter, the WD momentarily overflows its Roche lobe, i.e., $R_* \gtrsim R_{\rm lobe}$, and the instantaneous Roche lobe radius is
\begin{equation}
R_{\rm lobe} \simeq \beta_0 R_{\rm p} (M_{\rm h} / M_*)^{-1/3}.
\label{eq:Roche_lobe}
\end{equation}
An exterior layer of the WD is tidally stripped, where the exact value of $\beta_0$ depends on the rotation of WD, the orbital eccentricity, and the mass ratio $M_{\rm h} / M_*$ \citep{Sepinsky_Equipotential_2007}. Here we take $\beta_0 \simeq 0.5$.

We assume the stripped layer is a spherical shell at the surface, and its depth is 
\begin{equation}
z = R_* - R_{\rm lobe} \simeq \left( 1 - \frac{\beta_0}{\beta} \right) R_*,
\label{eq:z}
\end{equation}
then the stripped mass is 
\begin{equation}
\Delta M = 4 \pi R_*^2 \int^z_0 \rho(z') dz'.
\label{eq:dm}
\end{equation}
In Appendix \ref{subsec:LE} we re-derive the density structure of a WD, following \cite{Chandrasekhar_Stellar_1935}, which will also be used in our hydrodynamic simulation. The hydrostatic balance at the surface is given by $dP / dr = - G \rho M_* / R_*^2$, where $\rho$ is the density. Using Eq. (\ref{eq:dP_dr2}), we can obtain
\begin{equation}
\frac{1}{2} \frac{d(x_{\rm F}^2)}{dr} = -G\frac{\rho_0}{P_0} \frac{GM_*}{R_*^2}.
\label{eq:d1xf}
\end{equation}
Here $x_{\rm F} \equiv p_{\rm F} / (m_{\rm e} c)$ is the nomalised Fermi momentum; we have used $x_{\rm F} \ll 1$ for the non-relativistic gas on surface of the WD. The constants $P_0$ and $\rho_0$ are given by Eq. (\ref{eq:P0}) and (\ref{eq:rho0}), respectively. Using the relation $\rho = \rho_0 x_{\rm F}^3$ (Eq. \ref{eq:rho}), we can obtain from the above the density gradient at the WD surface
\begin{equation}
\frac{1}{2} \frac{d(\rho/\rho_0)^{2/3}}{dr} \simeq -G\frac{\rho_0}{P_0} \frac{GM_*}{R_*^2}.
\label{eq:drho_dr}
\end{equation}

Using $dz' = - dr$ and integrating Eq. (\ref{eq:drho_dr}) with respect to $z'$,  we can obtain the leading-term of density on the surface ($z' \ll R_*$)
\begin{equation}
\begin{split}
\rho(z') &\simeq \left( \frac{2G \rho_0^{5/3} M_*}{P_0 R_*} \right)^{3/2} \left(\frac{z'}{R_*}\right)^{3/2} \\
&\simeq 4.2 \left[1-(M_*/M_{\rm ch})^{4/3}\right]^{3/4} \bar{\rho}_* \left( \frac{z'}{R_*} \right)^{3/2}.
\end{split}
\label{eq:rho_z}
\end{equation}
Here the second equation is obtained by the mass-radius relation Eq. (\ref{eq:m_r}).

Plugging Eq. (\ref{eq:rho_z}) into Eq. (\ref{eq:dm}) and carrying out the integral, one can obtain the stripped mass with respect to $\beta$ as
\begin{equation}
\frac{\Delta M}{M_*} \simeq 4.8 \left[1-(M_*/M_{\rm ch})^{4/3}\right]^{3/4} \left( 1 - \frac{\beta_0}{\beta} \right)^{5/2}.
\label{eq:dm2}
\end{equation}
Later in Section 3, we perform the hydrodynamic simulations to verify this equation.

\cite{Zalamea_White_2010} estimated the stripped mass as $\Delta M / M_* \simeq 6.1 (1- M_*/M_{\rm ch})^{0.67} (1 - \beta_0/\beta)^{5/2}$, which We compare with Eq. (\ref{eq:dm2}) in Figure \ref{fig:dm_beta}, and find that they are close to each other. Note that \cite{Shen_Fast_2019} also estimated the stripped mass (their Eq. 14) by assuming that the shape of the stripped layer is a slice of a sphere rather than a spherical shell.


\subsection{Mass fallback}
\label{subsec:fall_back_rate}
After the tidal stripping, the most bound debris will return to pericenter at a time $t_{\rm fb}= 2\pi \sqrt{a_{\rm mb}^3 / (G M_{\rm h})}$, where $a_{\rm mb}$ is its semi-major axis, and the less bound debris follow it to fall back. For the eccentric orbit we consider here, the survived WD separates the debris stream into two arms. The front arm returns to the pericenter earlier, and the back arm returns later following the survived WD. Therefore, the mass fallback rate should have two peaks (see Figure \ref{fig:fall_back}), which are related to two arms of the stream.

Adopting the “frozen-in” model in TDEs \citep{Lodato_Stellar_2009}, in which the spread in specific energy is $\Delta \epsilon \simeq \xi G M _{\rm h}R_*/ R_{\rm p}^2$, the specific energy of the most bound debris is given by
\begin{equation}
\epsilon_{\rm mb} = -\frac{G M_{\rm h}}{2 a_{\rm mb}} \simeq -\frac{G M_{\rm h}}{2 a} - \frac{\xi G M_{\rm h} R_*}{R_{\rm p}^2}
\label{eq:E_mb}
\end{equation}
where we use a coefficient $\xi \gtrsim 1$ to quantify the uncertainty of the tidal effects on the WD, e.g., the tidal deformation and the tidal spin-up \citep{Rees_Tidal_1988,Goicovic_Hydrodynamical_2019}. Then the semi-major $a_{\rm mb}$ can be easily derived by Eq. (\ref{eq:E_mb}):
\begin{equation}
a_{\rm mb} \simeq \left(a^{-1} + \frac{2 \xi R_*}{R_{\rm p}^2}\right)^{-1}.
\label{eq:a_mb}
\end{equation}

Using the above equation, we can write $t_{\rm fb}$ in unit of $P$ as
\begin{equation}
t_{\rm fb} \simeq \left[1 + \frac{2 \xi \beta}{1-e} \left( \frac{M_{\rm h}}{M_*} \right)^{-1/3} \right]^{-3/2} P.
\label{eq:t_fb}
\end{equation}
We will compare this relation with the simulation result in Section \ref{subsec:Mfb_sim} (see Figure \ref{fig:ratio_tfb}).

The stripped material forms two arms of stream, each arm on one side of the WD. We can also estimate in what circumstance even the outer arm of the debris stream is bound to the IMBH. The specific energy of the tail end of the outer arm is
\begin{equation}
\epsilon_{\rm tail} \simeq -\frac{G M_{\rm h}}{2 a} + \frac{\xi G M_{\rm h} R_*}{R_{\rm p}^2}.
\label{eq:E_tail}
\end{equation}

If $\epsilon_{\rm tail} \lesssim 0$, i.e., 
\begin{equation}
1-e \gtrsim 0.078\ \xi \beta M_{\rm h,4}^{-1/3} \left(\frac{M_*}{0.6\ M_{\odot}} \right)^{1/3},
\label{eq:E_tail2} 
\end{equation}
then all of the debris are bound to the IMBH. It gives a critical orbital eccentricity $e_{\rm crit}$ for no unbound debris being produced. Note that \cite{Hayasaki_Finite_2013} considered eccentric-orbit TDEs and derived a similar $e_{\rm crit}$. However, because they studied a different parameter regime ($\beta \gtrsim 1$, i.e., for full TDEs), their expression for the specific energy has the $R_{\rm T}^{-2}$ dependence instead of $R_{\rm p}^{-2}$. Therefore, they derived $1 - e_{\rm crit} \propto \beta^{-1} (M_{\rm h}/M_*)^{-1/3}$, which is different from ours.

Another useful property is the peak time $t_{\rm peak}$ of the fallback rate, which depends on the density and the specific energy distribution in the stripped layer of the WD. Because the tidal deformation affects the shape of the stripped layer, it is hard to determine its value. The rise timescale of the fallback rate is $t_{\rm rise} = t_{\rm peak} - t_{\rm fb}$. If the radiation luminosity is proportional to the mass fallback rate $\dot M_{\rm fb}$, $t_{\rm rise}$ would be an important observable in an EM burst. In Section \ref{subsec:Mfb_sim}, we will use the simulation result to determine the ratio $t_{\rm rise}/t_{\rm fb}$ and study its dependences. It is $\sim 0.5$ -- $1$ for tidal disruption event with parabolic orbit $e = 1$ \citep{Guillochon_Hydrodynamical_2013}. However, it might be different for the elliptical orbit.

\section{Numerical simulation}
\label{sec:simulation}

We conduct the simulations of tidal stripping of a WD by an IMBH using FLASH (version 4.0), an adaptive-mesh grid-based hydrodynamics code \citep{Fryxell_FLASH_2000}. The Hydro implementation we adopt is the directionally split piecewise-parabolic approach \citep{Colella_PPM_1984} which is provided within the framework of the FLASH code. We adopt the modified gravity algorithm which is presented in \cite{Guillochon_Consequences_2011}, and the setting of the multipole gravity solver is the same as in \cite{Guillochon_Hydrodynamical_2013}. The simulations are preformed in the rest-frame of the WD.

The initial density profile of the WD is constructed by solving the equation of state of a degenerate star, which is presented in Appendix \ref{subsec:LE}. We set the IMBH mass to be $10^4\ M_{\odot}$ in our simulations, and separately run 9 simulations with different eccentricities $e = 0.7$, $0.8$, $0.9$ and impact factors $\beta = 0.55$, $0.6$, $0.7$, for two WD masses $M_* = 0.67$ and $1.07\ M_{\odot}$, respectively.

Note that the relativistic effects, e.g., apsidal procession and GW radiation, are not included in the simulations, thus, the WD's orbit is Keplerian. Here we mainly aim to study the tidal stripping near the pericenter in a single orbit with different orbital parameters; we do not study the orbital change due to GW radiation by simulations. Therefore, Keplerian motion should be sufficient for our purpose.

\begin{figure}		
\centering
 \includegraphics[scale=0.37]{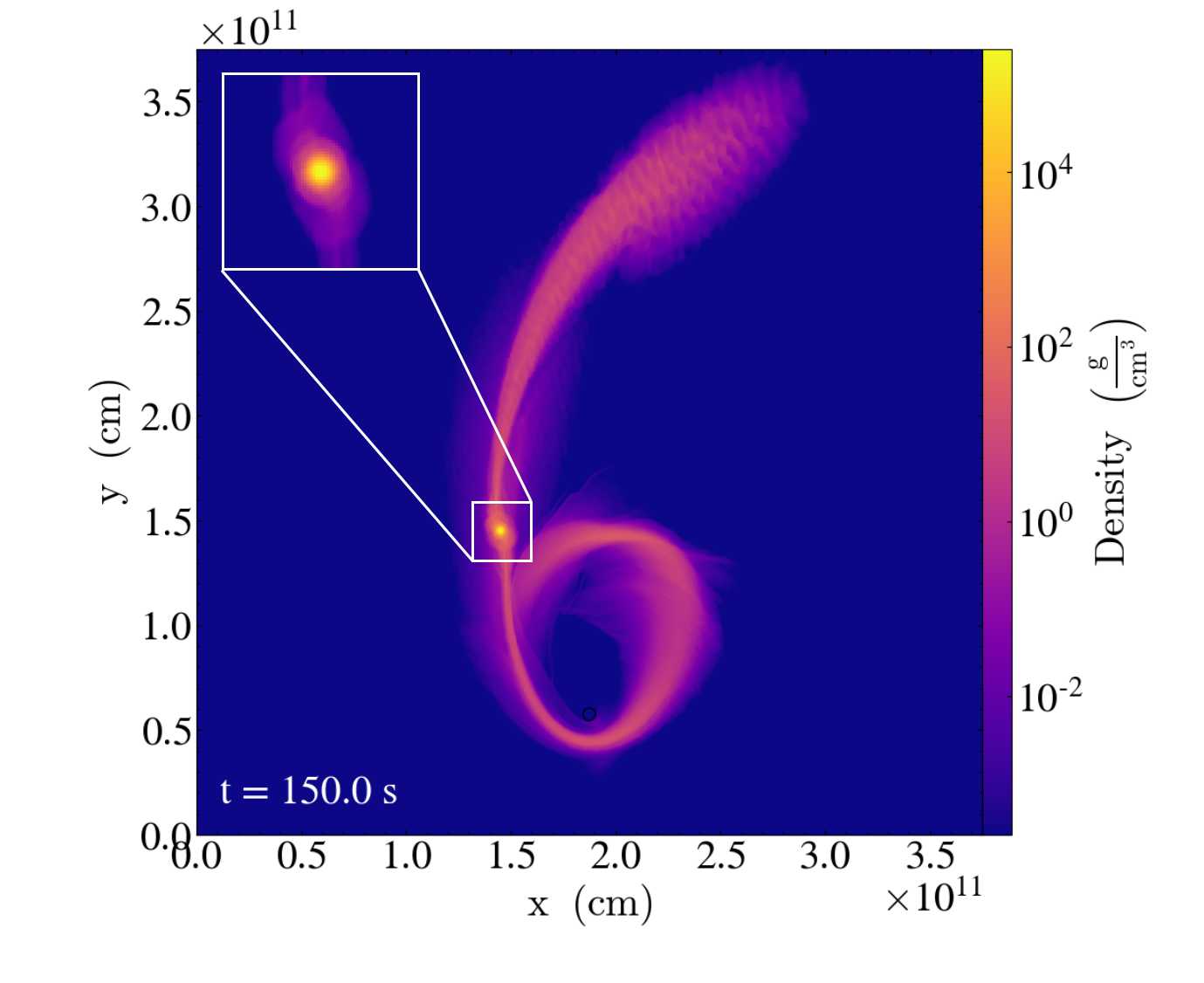}
\caption{Projection of the gas density on the orbital plane of tidal stripping of a $1.07\ M_{\odot}$ WD by a $10^4\ M_{\odot}$ IMBH with $e=0.8$ and $\beta=0.7$. The first returning debris have passed through the pericenter and form a ring-like structure. The WD which is shown in the inset diagram is returning to the pericenter. The simulation is conducted by the FLASH hydrodynamics code. The time since the beginning of the simulation is shown in the lower left. The IMBH is indicated by the black circle.}
\label{fig:snapshot}
\end{figure}

In the beginning of the simulation, we place the WD at a distance of $5 R_{\rm T}$ from the central IMBH and relax it for $50$ s, which is $\sim 25$ times of its sound crossing timescale. The size of the domain is $5 \times 10^{12}$ and $1.5 \times 10^{12}$ cm for WD mass $0.67$ and $1.07\ M_{\odot}$, respectively. The domain is initially composed of $8^3$ blocks, which is then refined to be the smaller blocks in accordance with the gas density. The maximum level of refinement is $16$, corresponding to a minimum cell size of $\sim 1.9 \times 10^7$ and $\sim 5.7 \times 10^6$ cm ($\sim 0.02$ and $\sim 0.01\ R_*$) for WD mass $0.67$ and $1.07\ M_{\odot}$, respectively.

After being tidally stripped by the IMBH, the survived WD will return to the pericenter with the stretched debris stream on the two sides. Figure \ref{fig:snapshot} provides a snapshot, where the most bound debris have passed through the pericenter, and the debris tail follows behind the WD. The goal here is to study the orbital dependences of the stripped mass and the mass fallback rate.


\subsection{Dependences of the stripped mass}
\label{subsec:dM_sim}
Determining how much mass is being stripped away and how much remains to the survived WD is not straightforward, due to the dynamical nature of the process where the separation between the WD and the BH changes rapidly. We adopt the iterative approach in \cite{Guillochon_Hydrodynamical_2013} to calculate the mass that is bound to the WD after the stripping. The total stripped mass is equal to the initial WD mass minus the bound mass. The stripped mass is calculated when the WD is at the apocenter.

\begin{figure}		
\centering
 \includegraphics[scale=0.5]{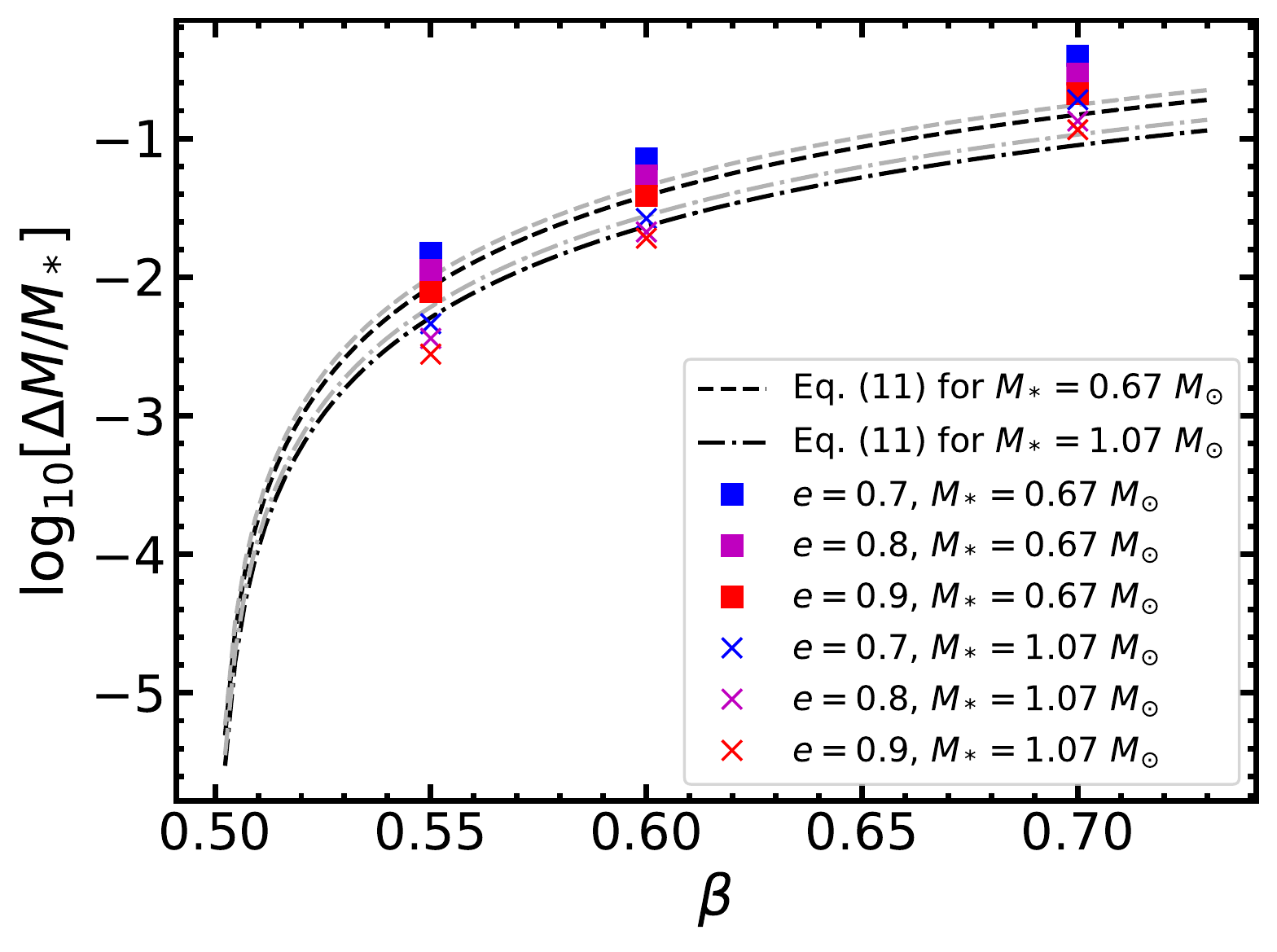}
\caption{The relationship between the total stripped mass of a WD and impact factor $\beta$. The data is the simulation result with different parameters (eccentricity $e$ and WD mass $M_*$). The black dashed and dotted-dashed lines represent the analytical result (Eq. \ref{eq:dm2}) for two different WD masses $0.67$ and $1.07\ M_{\odot}$, respectively. In general, the simulation results are consistent with the analytical calculations for $\beta \lesssim 0.7$. For comparison, we also plot the analytical result of \cite{Zalamea_White_2010} with the gray lines.}
\label{fig:dm_beta}
\end{figure}

The simulation results and the analytical results of the $\Delta M - \beta$ relation are compared in Figure \ref{fig:dm_beta}. In general, Eq. (\ref{eq:dm2}) can well represent the $\Delta M - \beta$ relation, except for the large impact factors $\beta \gtrsim 0.7$, where Eq. (\ref{eq:dm2}) slightly underestimates $\Delta M$. This is probably due to the fact that at large $\beta$, the WD is tidaly stretched near $R_{\rm p}$, so more mass moves out of the Roche lobe than that given by Eqs. (\ref{eq:z} -- \ref{eq:rho_z}). However, for $\beta < 0.7$, these equations are applicable and so is Eq. (\ref{eq:dm2}).

Figure \ref{fig:dm_beta} shows that the orbital eccentricity slightly affects the stripped mass: the stripped mass is a little larger for the smaller eccentricity. This might be due to the fact that the tidal force can act on the WD much longer near the pericenter if the eccentricity is smaller, and the mass on the surface would have more time to be stripped away from the WD. However, we do not intent to quantify this effect in the analytical calculation.

It would be ideal to check the consistency between the analytical result Eq. (\ref{eq:dm2}) and the numerical result in a wider dynamical range. Unfortunately, for small impact factors $\beta \sim 0.5$, the stripped mass is too low to be accurately calculated in the simulations due to the lack of resolution.

\subsection{Mass Fallback Rate}
\label{subsec:Mfb_sim}

The mass fallback rate is determined by the distribution of the specific binding energy within the stripped mass $dM/dE$, i.e., $\dot M_{\rm fb}(t) = (dM/dE) \times (dE/dt)$, where $dE/dt$ is given by the Kepler's third law. This method is valid when the debris is stretched to be a long stream so each part of the material in the stream can freely move on its own orbit.

\begin{figure*}		
\centering
\begin{minipage}[t]{0.48\textwidth}
\centering
 \includegraphics[scale=0.27]{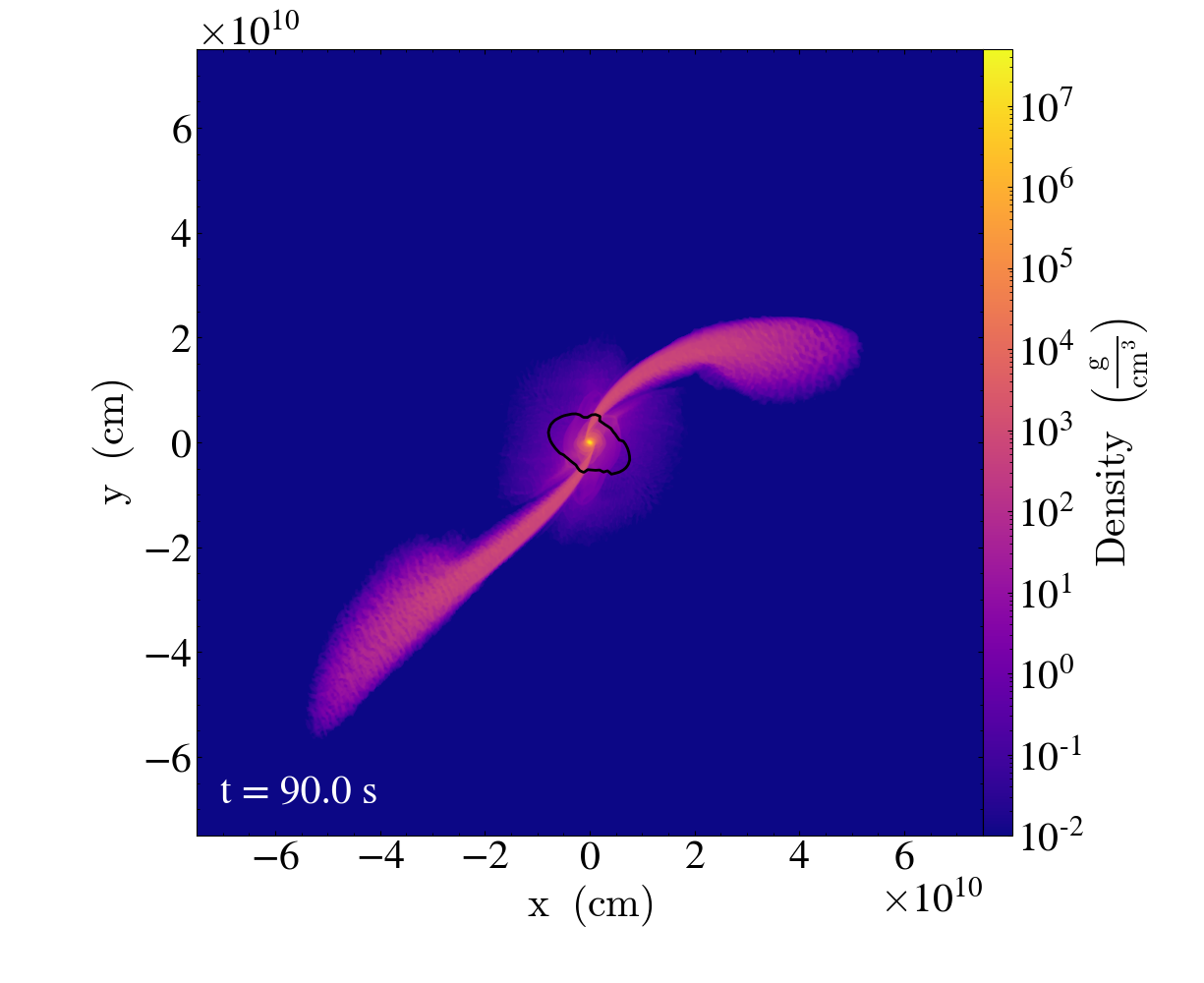}
 \end{minipage}
 \begin{minipage}[t]{0.48\textwidth}
 \centering
 \includegraphics[scale=0.27]{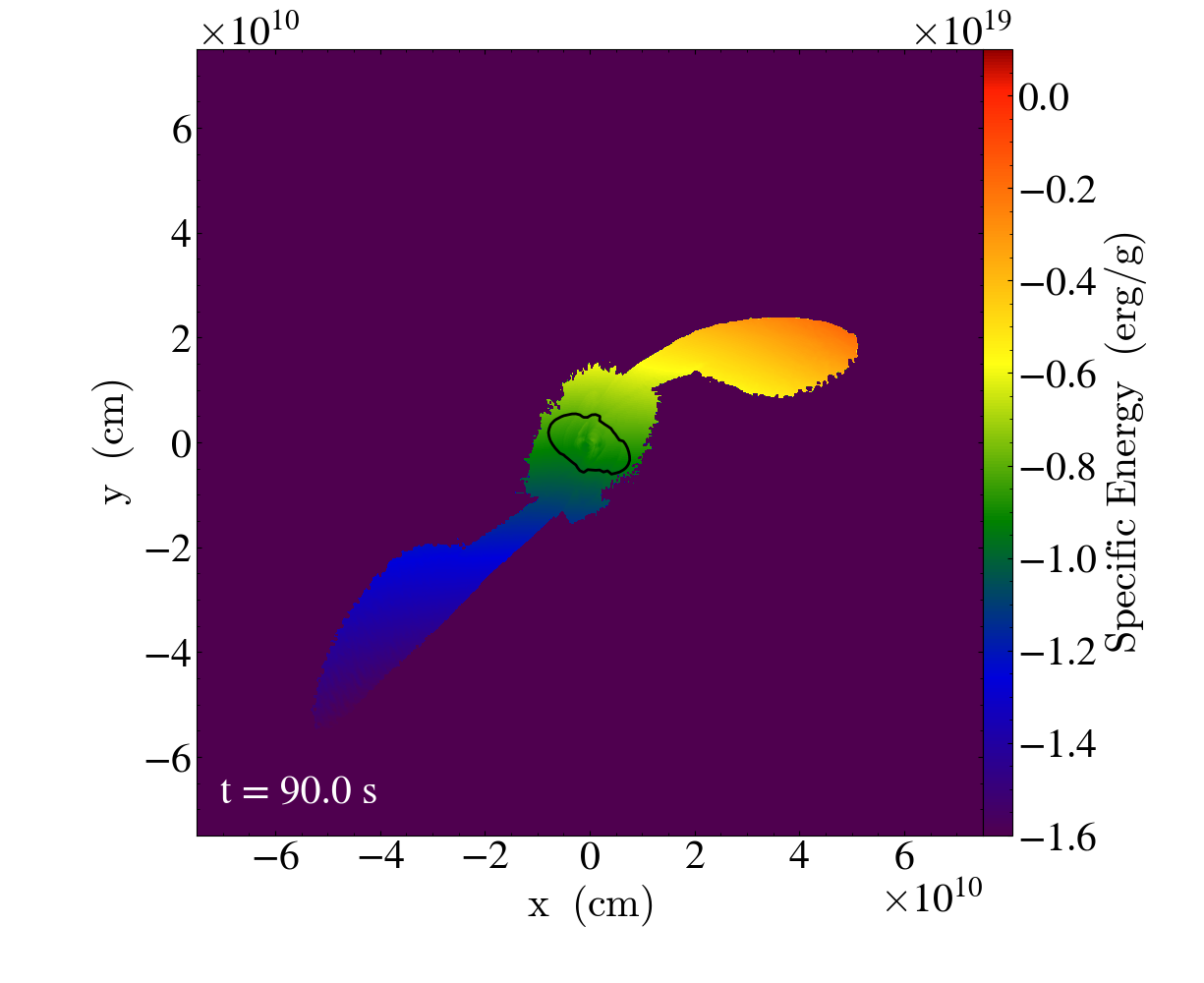}
  \end{minipage}
\caption{Snapshot of tidal stripping generated from the same simulation of Figure \ref{fig:snapshot} at different time. The left and right panels show the projection of the density and the specific energy distribution on the orbital plane, respectively. The material enclosed in the black line is bound to the WD. The debris stream on the left side is more bound to the IMBH than the one on the right side; thus, it will firstly return to the pericenter. Furthermore, the specific energy of the material are all minus, therefore, these material are all bound to the IMBH. The mass fallback rate, which is shown in Figure \ref{fig:fall_back_WD}, is generated from this snapshot.}
\label{fig:snapshot_e}
\end{figure*}

\begin{figure*}		
\centering
\begin{minipage}[t]{0.48\textwidth}
\centering
 \includegraphics[scale=0.48]{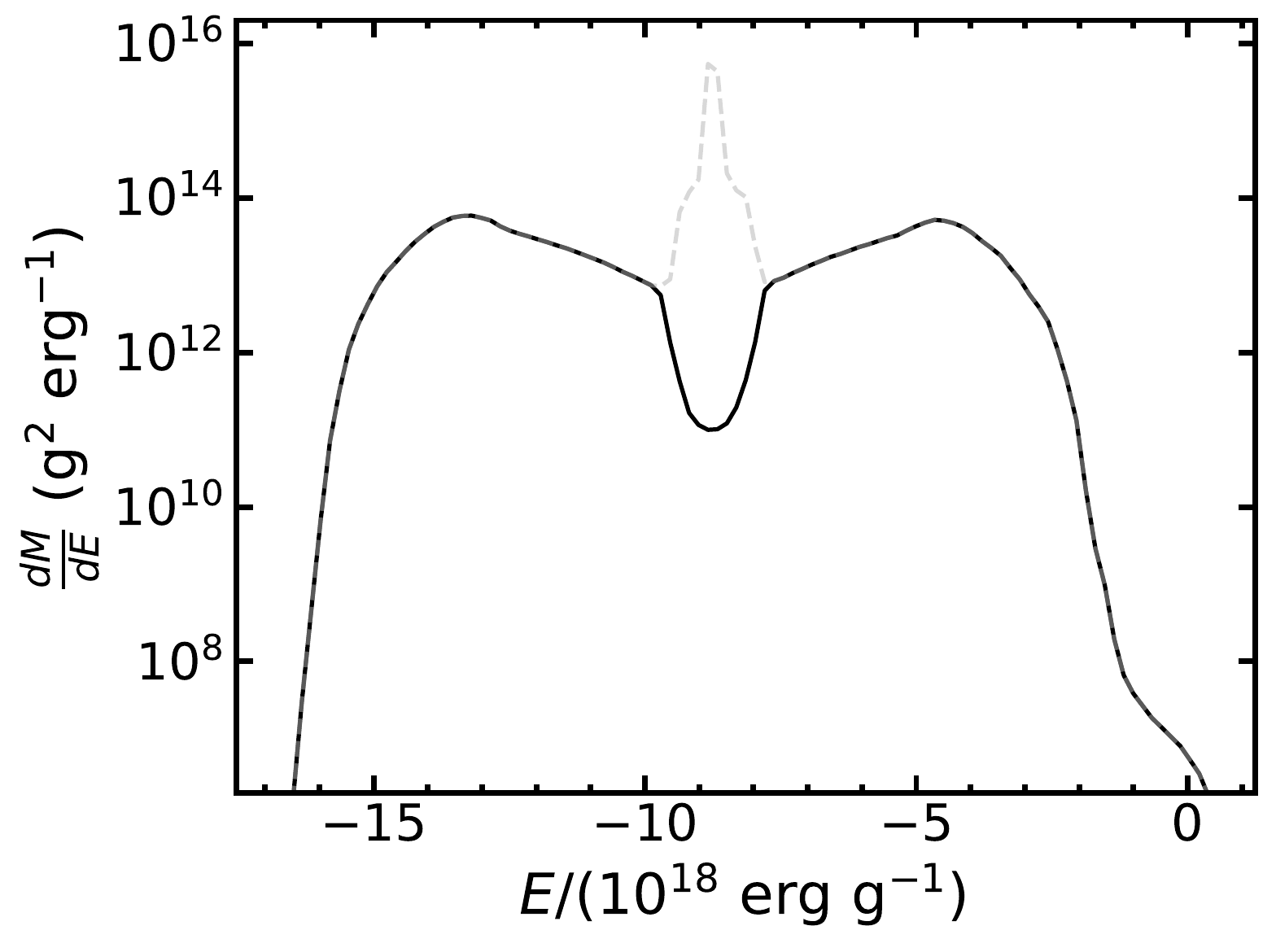}
 \end{minipage}
 \begin{minipage}[t]{0.48\textwidth}
 \centering
 \includegraphics[scale=0.48]{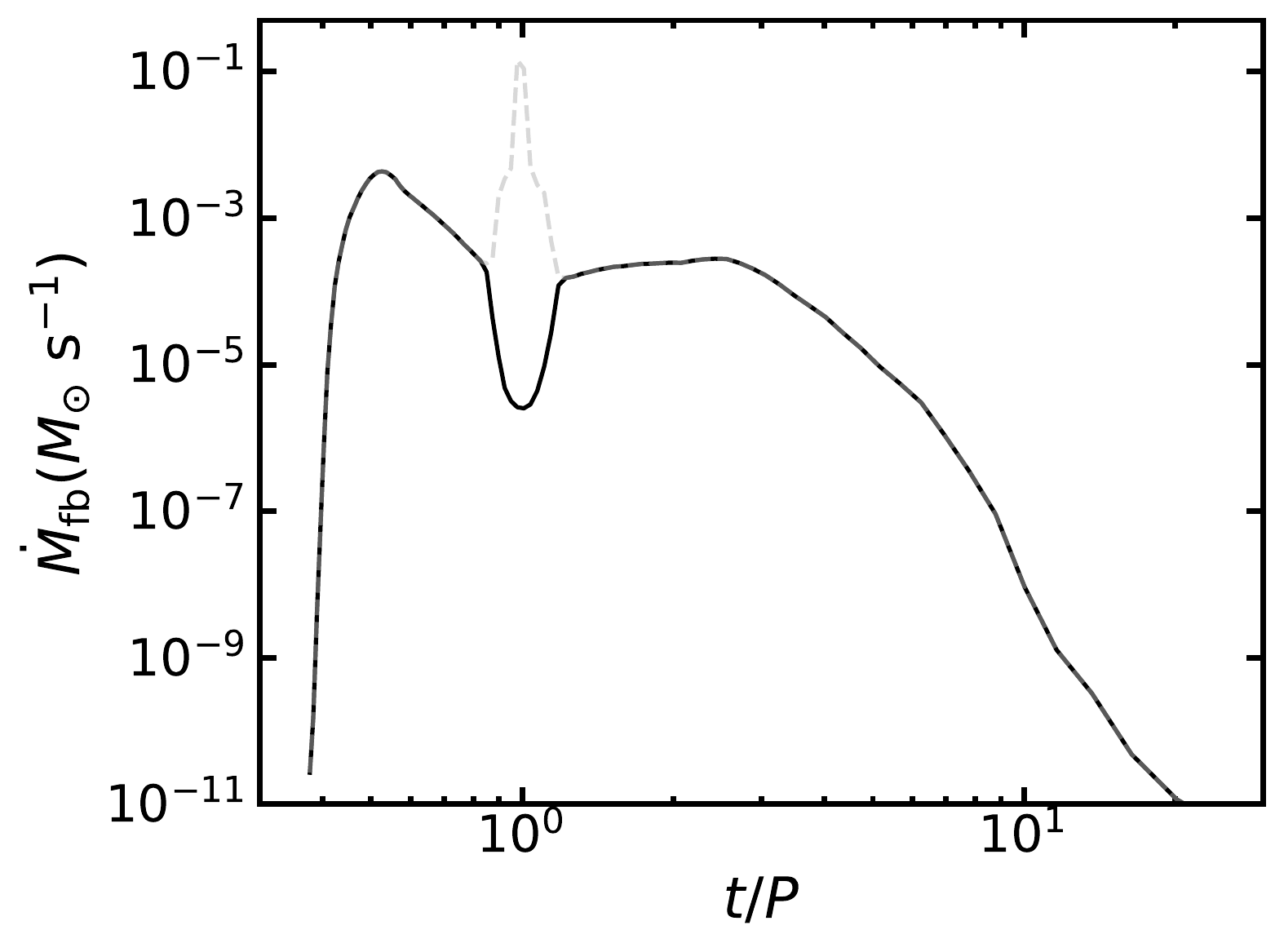}
  \end{minipage}
\caption{Distribution of specific energies (left) and evolution of mass fallback rate (right) for $e = 0.8$, $\beta = 0.7$, $M_{\rm h} = 10^4\ M_{\odot}$, and $M_* = 1.07\ M_{\odot}$ case, which are generated from the snapshot in Figure \ref{fig:snapshot_e}. The solid line is the result that has the contribution from the mass bound to the WD removed, and this contribution is shown by the gray dashed line. The WD mass surviving the tidal stripping would not supply to the accretion disk, but is subject to the next tidal stripping when it returns to the pericenter.}
\label{fig:fall_back_WD}
\end{figure*}

Figure \ref{fig:snapshot_e} shows a snapshot that we use to generate the mass fallback rate for the case of $e= 0.8$, $\beta = 0.7$, $M_* = 1.07\ M_{\odot}$. The left and right panels show the density and specific energy distribution, respectively. The material enclosed by the black line, which are bound to the WD, are removed in the calculation of the mass fallback rate, as is shown by the troughs in the $dM/dE$ distribution and the fallback rate curve, respectively, in Figure \ref{fig:fall_back_WD}. The gray dashed lines in Figure \ref{fig:fall_back_WD} represent the result without the removal of the bound mass.

We suggest that the stripped mass will be accreted when it returns and supplies to the pre-existing or newly-formed accretion disk. The material bound to the WD will not supply to the accretion disk but will be subject to the next tidal stripping when it returns to the pericenter. Therefore, we remove the mass bound to the WD in the calculation of fallback rate.

\begin{figure*}		
\centering
\begin{minipage}[t]{0.48\textwidth}
\centering
 \includegraphics[scale=0.48]{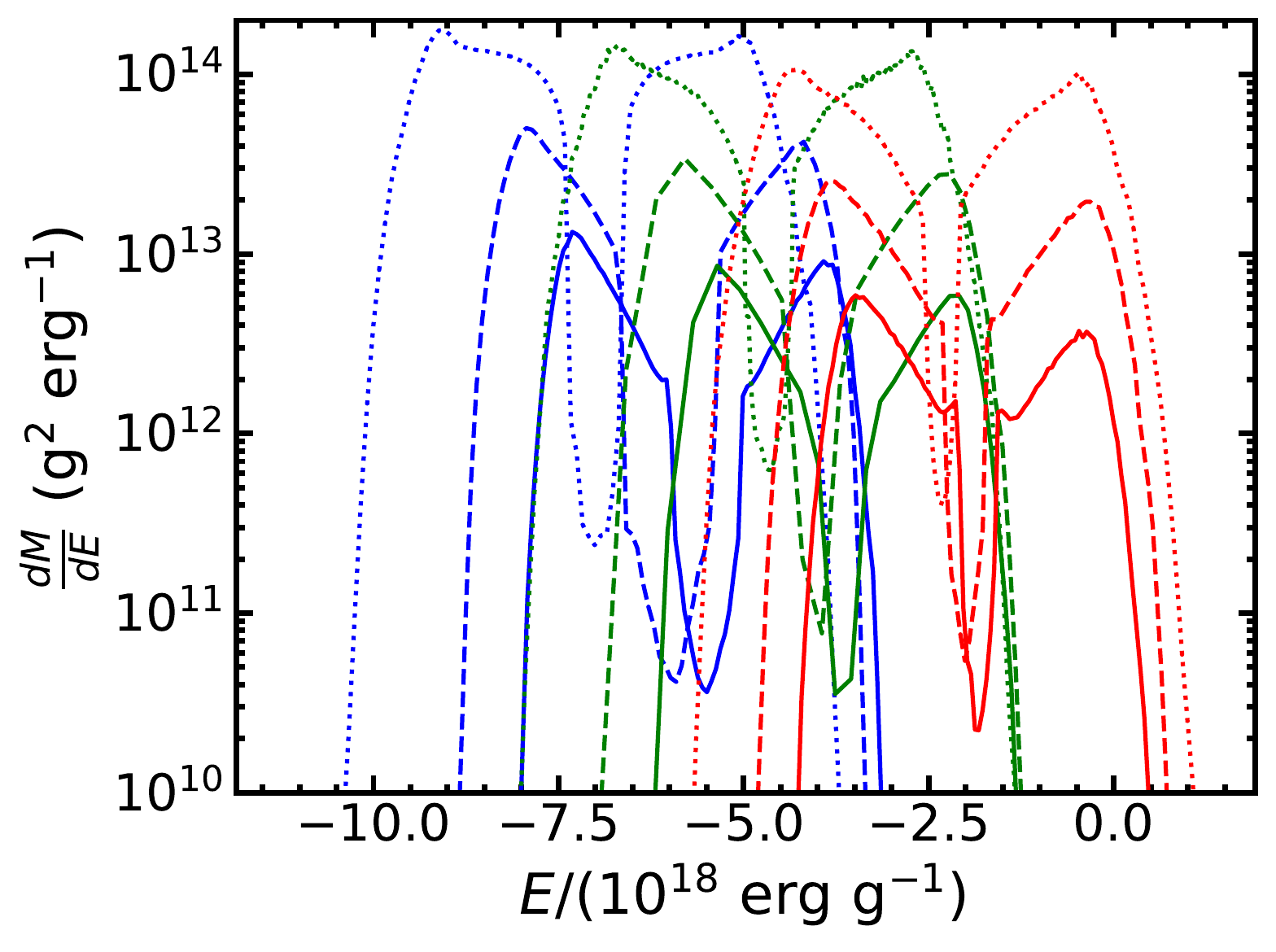}
 \end{minipage}
 \begin{minipage}[t]{0.48\textwidth}
 \centering
 \includegraphics[scale=0.48]{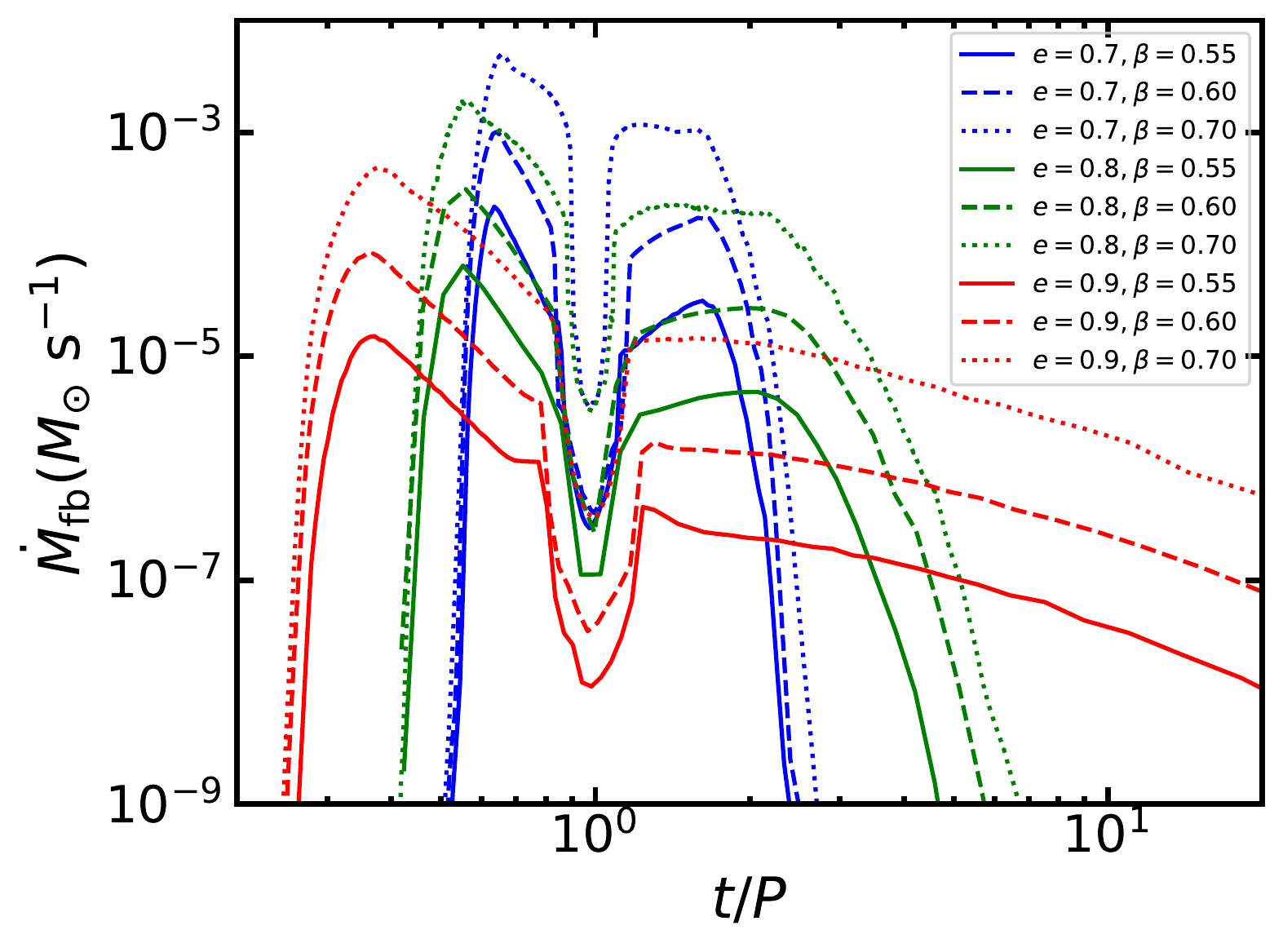}
  \end{minipage}
\caption{Distribution of specific energies (left) and evolution of mass fallback rate (right) for the WD mass $M_* = 0.67\ M_{\odot}$ with different orbital parameters ($e$ and $\beta$).}
\label{fig:fall_back}
\end{figure*}

\begin{figure*}		
\centering
\begin{minipage}[t]{0.48\textwidth}
\centering
 \includegraphics[scale=0.48]{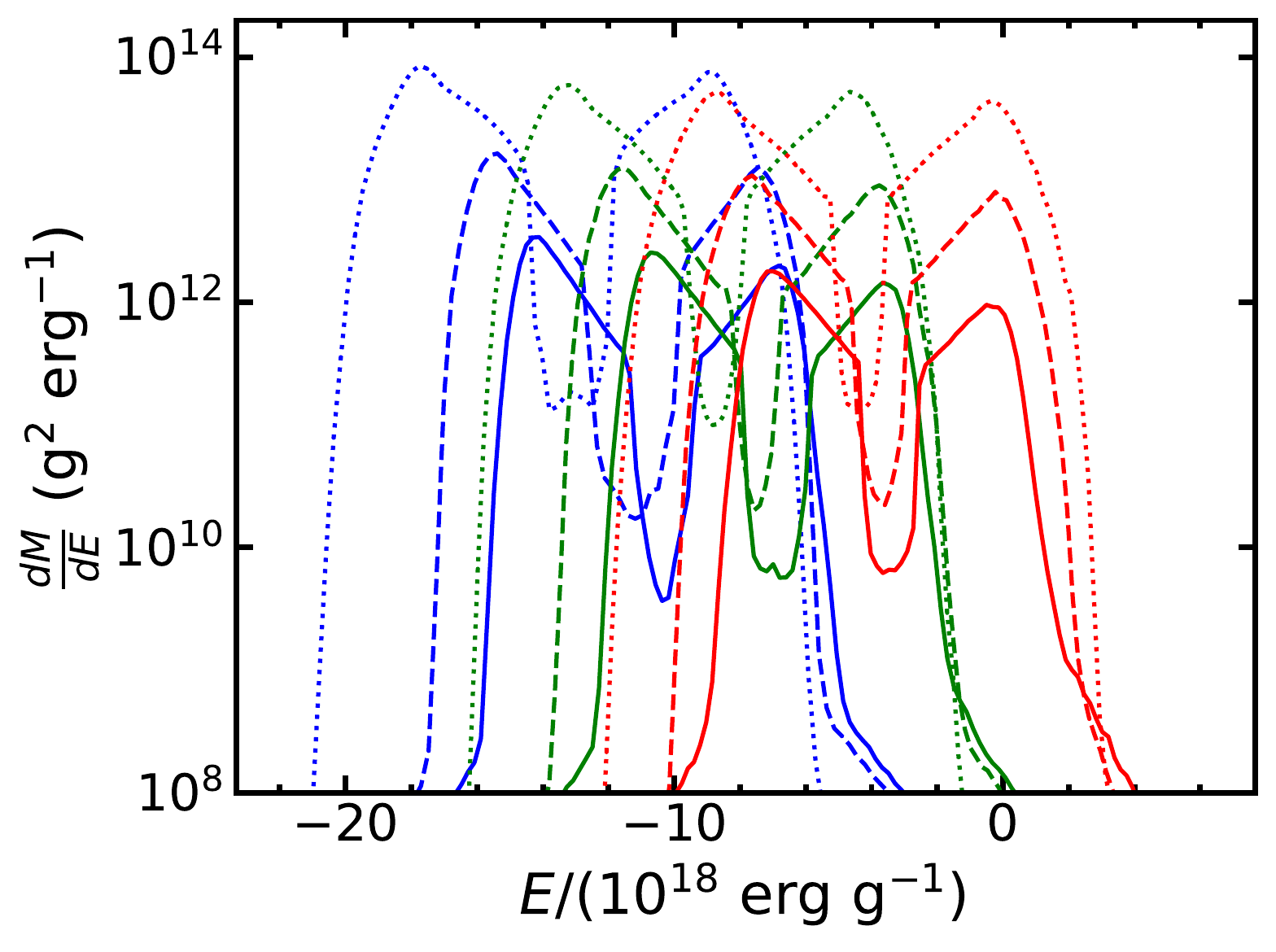}
 \end{minipage}
 \begin{minipage}[t]{0.48\textwidth}
 \centering
 \includegraphics[scale=0.48]{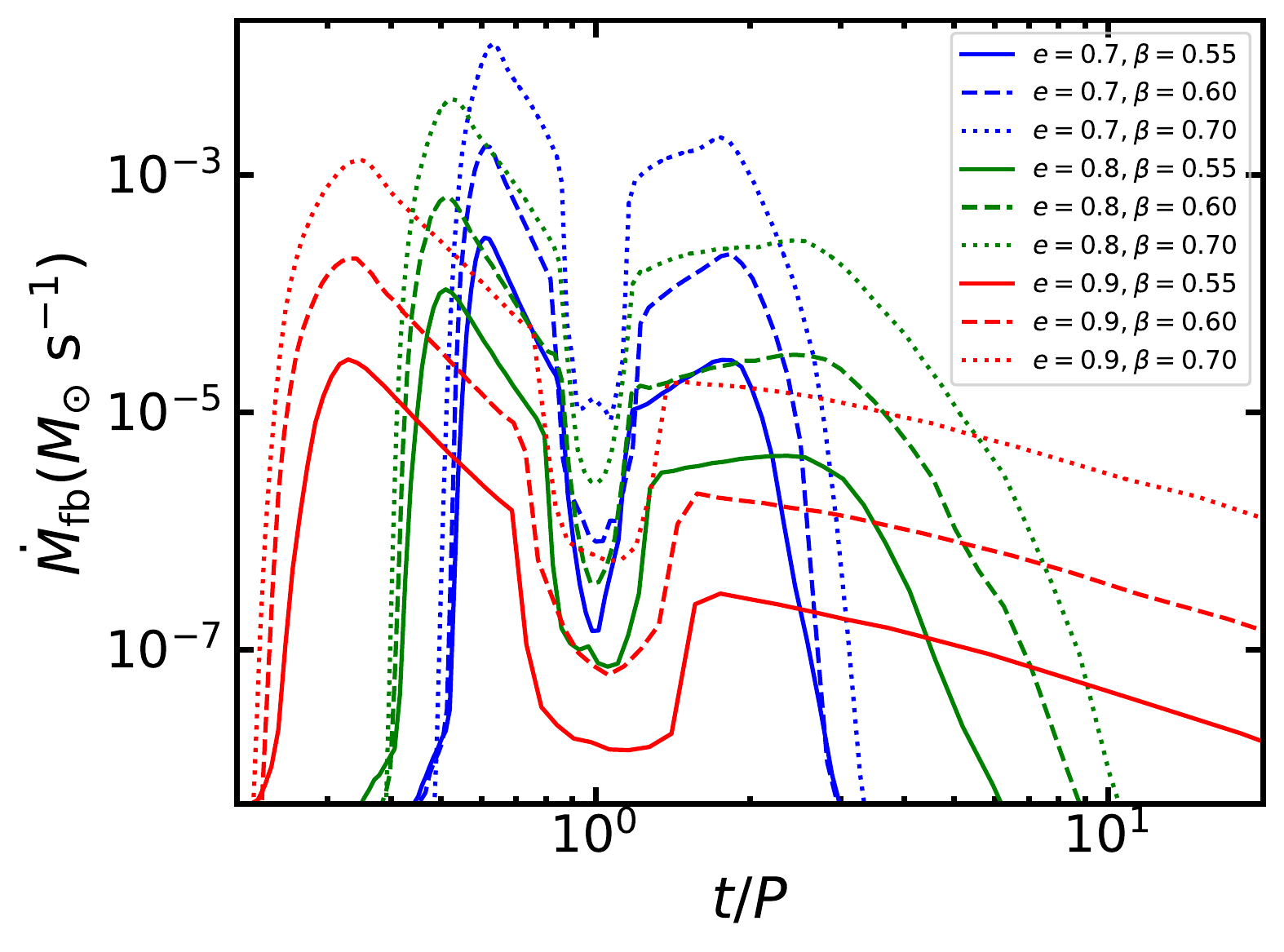}
  \end{minipage}
\caption{Same as Figure \ref{fig:fall_back} but for $1.07\ M_{\odot}$ WD mass.}
\label{fig:fall_back_5e7}
\end{figure*}

Our simulation results for $0.67$ and $1.07\ M_{\odot}$ WDs are plotted in Figures \ref{fig:fall_back} and \ref{fig:fall_back_5e7}, respectively. The left peak and the right peak of the valley-shaped fallback rate curve correspond the front arm and the back arm of the debris stream, respectively. The front arm returns to the pericenter earlier and the back arm returns following the WD. The left peak of the fallback rate is more than one order of magnitude higher than the right peak. We find that the total fallback mass of the left peak is slightly larger than that of the right peak by a factor of $1.2$ --  $1.5$. Therefore, the accretion of the first returned mass should dominate over the latter accretion of the stream tail.

Our simulation result can constrain $\xi$, the parameter that accounts for the tidal deformation and spin-up effects at the pericenter passage (Eq. \ref{eq:E_mb}). For the $M_* = 0.67\ M_{\odot}$ case, Figure \ref{fig:fall_back} left shows that all of the stripped mass are bound to the IMBH ($E < 0$) for $e = 0.7$ and $0.8$. However, for the large eccentricity ($e = 0.9$), there is a small amount of the mass that is unbound. Therefore, the critical eccentricity is $e_{\rm crit} \simeq 0.8$ -- $0.9$. Using Eq. (\ref{eq:E_tail2}) we can obtain $\xi \simeq 2$ -- $4$. Similarly, we can obtain the critical eccentricity $e_{\rm crit} \simeq 0.7$ -- $0.8$, and thus $\xi \simeq 3.5$ -- $5$ for the $1.07\ M_{\odot}$ WD case. Although the determination of the exact location of the stream tail is not firm, these estimates can constrain $\xi$ and give a physical interpretation of the condition for producing no unbound mass in a tidal stripping. In the following, we see that the value of $\xi$ can be further constrained by comparing the simulation results with the analytical calculation of $t_{\rm fb}$.

\begin{figure}		
\centering
 \includegraphics[scale=0.5]{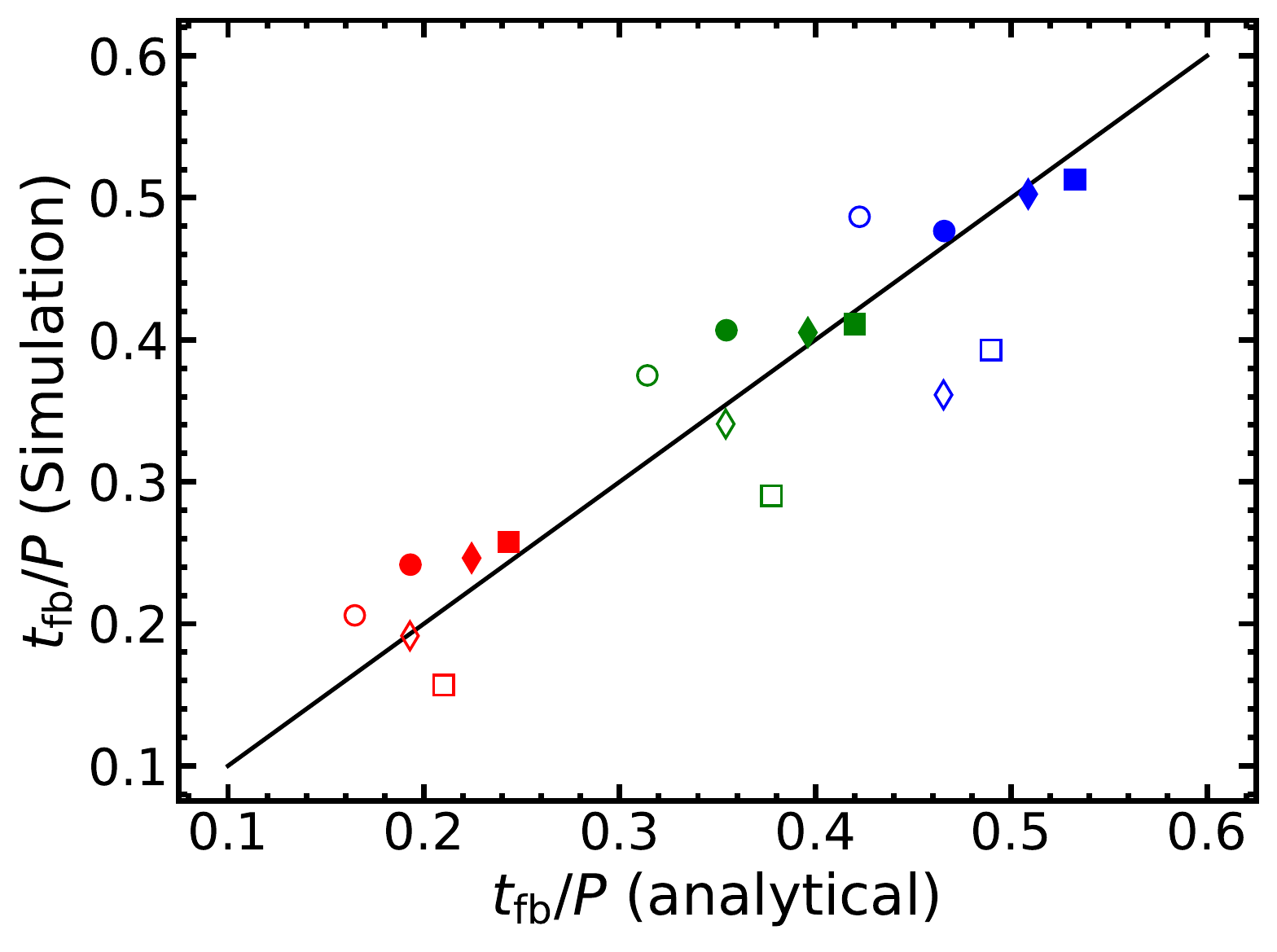}
\caption{Ratio between the fallback timescale $t_{\rm fb}$ and the orbital period $P$ of the WD. The x-axis is the analytical result of $t_{\rm fb}/P$ given by Eq. (\ref{eq:t_fb}), and y-axis is the simulation result of $t_{\rm fb}/P$. The solid line represents $y=x$. Solid and hollow points represent $M_* = 0.67$ and $1.07\ M_{\odot}$ WD masses, respectively. Square, diamond, and circular shapes represent $e = 0.7$, $0.8$, and $0.9$, respectively. Purple, green, and red colors represent $\beta = 0.55$, $0.6$, and $0.7$, respectively. The simulation results are well consistent with the analytical calculation with $\xi = 3.5$ as is shown by the black solid line.}
\label{fig:ratio_tfb}
\end{figure}

The larger the orbital eccentricity $e$ is, the earlier the mass returns to pericenter ahead of the WD, as shown by Eq. (\ref{eq:t_fb}) and in Figures \ref{fig:fall_back}--\ref{fig:fall_back_5e7}. We compare the simulation result of $t_{\rm fb} / P$ with the analytical calculation Eq. (\ref{eq:t_fb}) in Figure \ref{fig:ratio_tfb}. We find that the simulation results overall are consistent with the analytical result with $\xi = 3.5$, though it is a crude estimate due to the fact that the exact beginning of the mass fallback is not firmly determined. 

\begin{figure}		
\centering
 \includegraphics[scale=0.5]{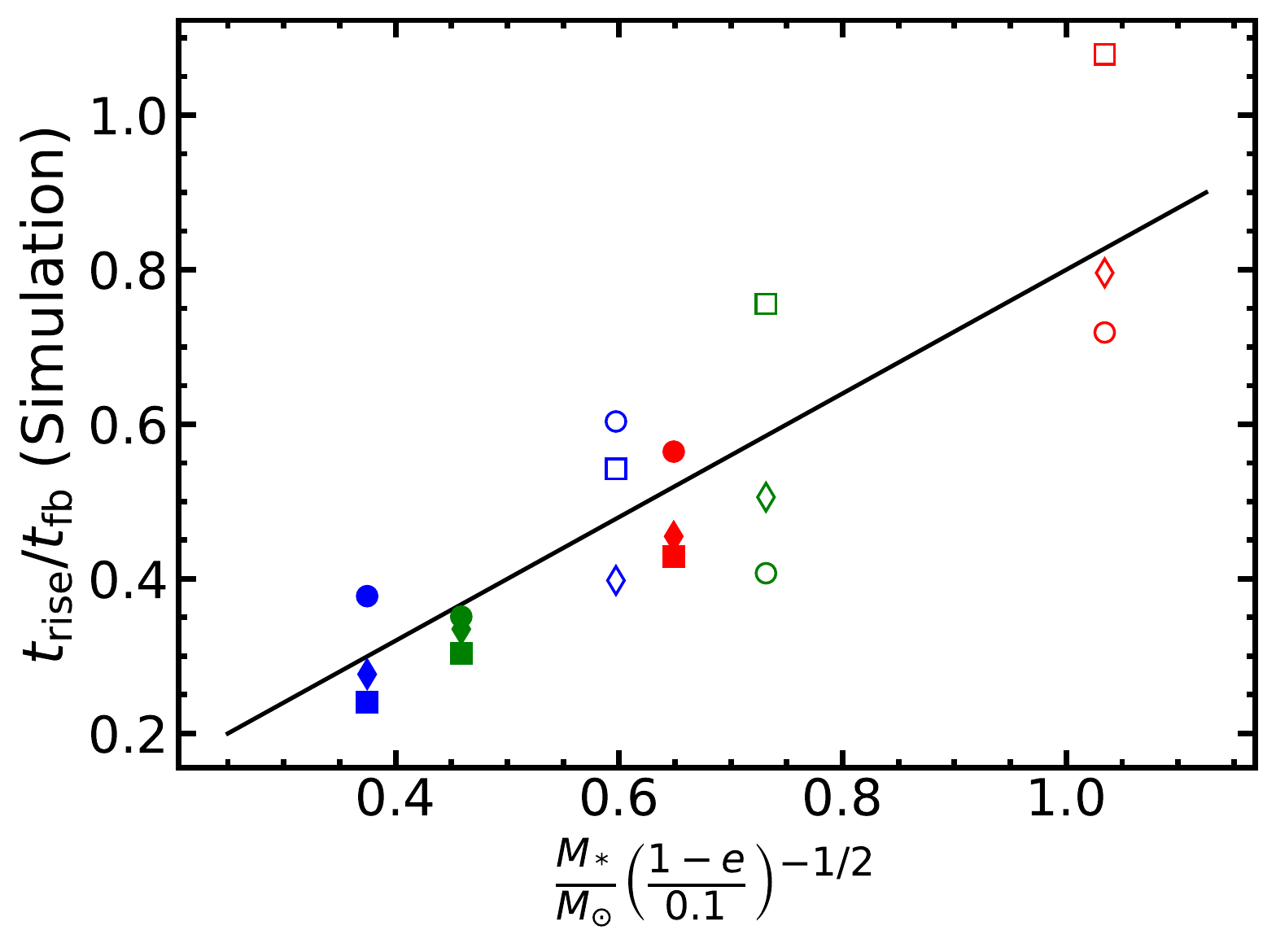}
\caption{Ratio between the rise timescale $t_{\rm rise} = t_{\rm peak} - t_{\rm fb}$ and the fallback timescale $t_{\rm fb}$. The fitting of the simulation result gives $t_{\rm rise}/t_{\rm fb} \simeq 0.8\ (M_*/M_{\odot}) [(1-e)/0.1]^{-1/2}$ represented by the solid line, however, we do not have a robust reason for this relation. Solid and hollow points represent $M_* = 0.67$ and $1.07\ M_{\odot}$ WD mass, respectively. Square, diamond, and circular shapes represent $e = 0.7$, $0.8$, and $0.9$, respectively. Purple, green, and red colors represent $\beta = 0.55$, $0.6$, and $0.7$, respectively.  The solid line shows the fitting result.}
\label{fig:tpeak_tfb}
\end{figure}

In Figure \ref{fig:tpeak_tfb}, we show the time scale ratio $t_{\rm rise}/t_{\rm fb}$ from the numerical result. We find that it can be described by an empirical relation
\begin{equation}
\frac{t_{\rm rise}}{t_{\rm fb}} \simeq 0.8\ \frac{M_*}{M_{\odot}} \left(\frac{1-e}{0.1}\right)^{-1/2}.
\label{eq:Gamma2_fit}
\end{equation}
Obviously, this relation cannot extent to $e \sim 1$ case; it is only valid for the intermediate eccentricity $e \sim 0.7$ -- $0.9$ that we consider here. Because we fix the BH mass to be $10^4\ M_{\odot}$ in the simulations, we cannot investigate the dependence of  $t_{\rm rise}/t_{\rm fb}$ on the BH mass.

\section{Evolution of the system}
\label{sec:evolution}
In this section we analytically calculate the long-term evolution of the system driven mainly by the GW radiation, until the WD is fully disrupted. The WD's radius increase due to its mass loss on each orbit is included.

The mean changing rate of semi-major axis and eccentricity due to GW radiation, averaged over one orbital period, are \citep{Peters_Gravitational_1964}
\begin{equation}
\dot a \simeq -\frac{64}{5} \frac{G^3 M_* M_{\rm h}^2}{c^5 a^3 (1-e^2)^{7/2}} \left(1+\frac{73}{24}e^2+\frac{37}{96}e^4\right),
\label{eq:da_dt}
\end{equation}
and
\begin{equation}
\dot e \simeq -\frac{304}{15} e \frac{G^3 M_* M_{\rm h}^2}{c^5 a^4 (1-e^2)^{5/2}} \left(1+\frac{121}{304}e^2\right),
\label{eq:de_dt}
\end{equation}
respectively. Their corresponding changes in one orbit are $P \dot a$ and $P \dot e$, respectively.

The average change rate of pericenter radius can be obtained from Eq. (\ref {eq:da_dt}) and Eq. (\ref {eq:de_dt}) as,
\begin{equation}
\begin{split}
\dot R_{\rm p} &= (1-e) \dot a - a \dot e \\
&= -\frac{64}{5} (1-e)^2 \frac{G^3 M_* M_{\rm h}^2}{c^5 a^3 (1-e^2)^{7/2}} \\
&\times \left(1-\frac{7}{12}e+\frac{7}{8}e^2+\frac{47}{192}e^3 \right).
\end{split}
\label{eq:dRp_dt}
\end{equation}

The WD's orbit shrinks and is circularized slowly due to the GW radiation. After its pericenter passes through the critical value ($\beta \gtrsim \beta_0$), the mass-loss begins. Define $R_0$, $M_0$, $R_{\rm p0}$, $a_0$, and $e_0$ to be the WD's radius and mass, the pericenter radius, the semi-major axis, and the eccentricity when $\beta=\beta_0=0.5$, respectively. We define a timescale on which the WD's orbit shrinks via GW radiation as 
\begin{equation}
\begin{split}
t_{\rm GW} &\equiv \frac{R_{\rm p0}}{|\dot R_{\rm p}|_{t=0}} \\
&\simeq 300 \, M_{\rm h,4}^{-2/3} \left(\frac{M_0}{0.6\ M_{\odot}}\right)^{-11/3} \left(\frac{1-e_0}{0.1}\right)^{-3/2} \ {\rm yr}.
\end{split}
\label{eq:t_GW}
\end{equation}
Here we have approximated the WD mass-radius relation to $R_* \simeq 9 \times 10^8\ (M_*/M_{\odot})^{-1/3}$ cm.

\begin{figure}		
\centering
 \includegraphics[scale=0.5]{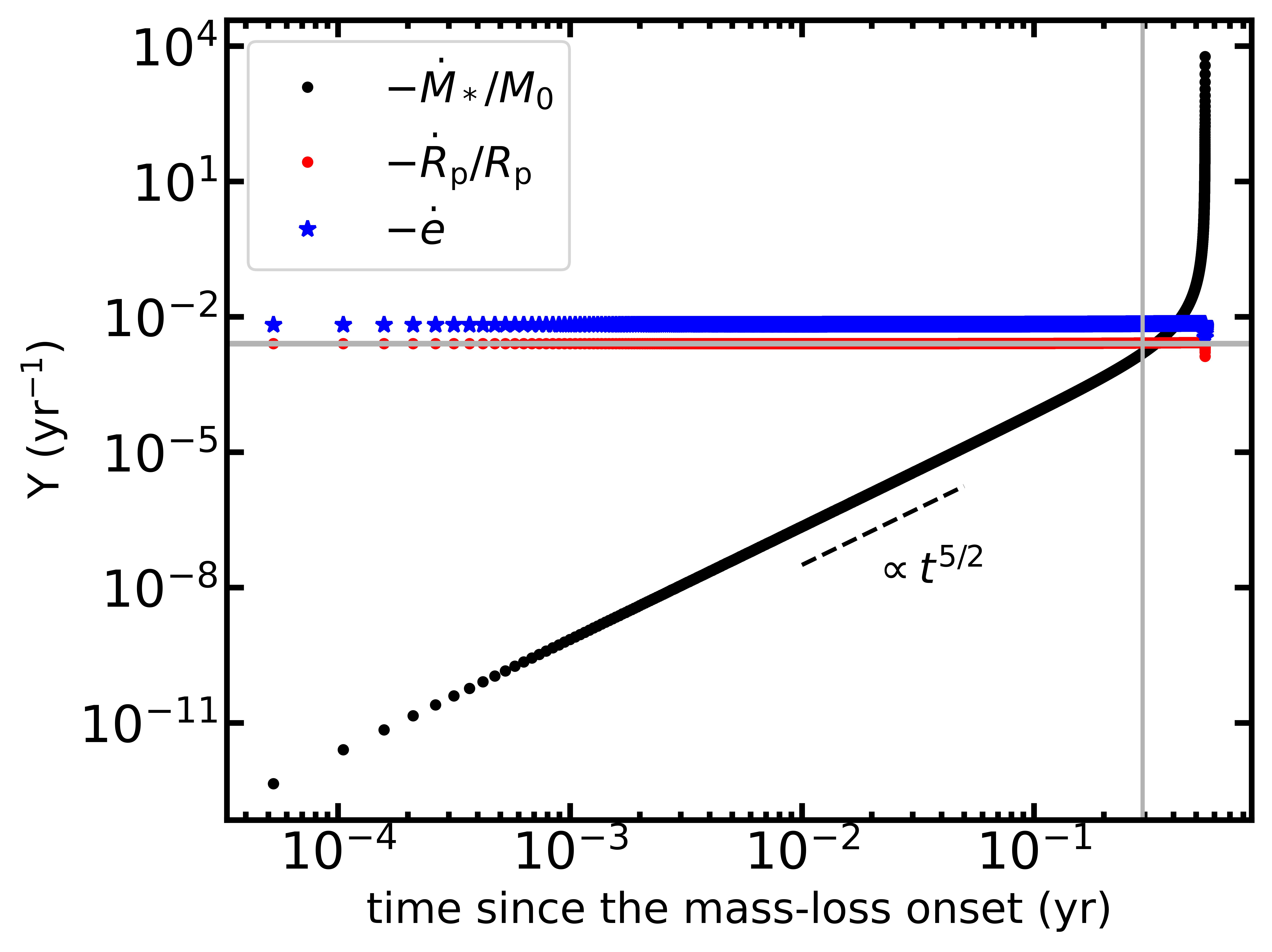}
\caption{Mass-loss evolution until the full disruption for tidal stripping of a $0.6\ M_{\odot}$ WD by an $10^4\ M_{\odot}$ IMBH with $e=0.9$. Y axis denotes the the orbit-averaged mass-loss rate $\dot M_* = -\Delta M / P$, the orbital change rate $-\dot R_{\rm p}/R_{\rm p}$ and $-\dot e$, calculated from Eq. (\ref{eq:da_dt}-\ref{eq:dm_evol}). At early time, the mass loss is mainly driven by the orbital shrinkage due to GW radiation, and $\dot M_* \propto t^{5/2}$. After a transitional timescale $t_{\rm tran}$ (represented by the gray vertical line), the WD becomes easier to be tidally stripped due to its expansion, thus, it will experience runaway mass loss right before the final disruption. The gray horizontal line represents $ Y= t_{\rm GW}^{-1}$.}
\label{fig:evolution}
\end{figure}

As the orbit evolves, we consider the successive tidal stripping in a time-dependent manner. Because $\beta = R_{\rm T}/R_{\rm p} \propto M_*^{-1/3} R_*/R_{\rm p}$, then using the WD mass-radius relation Eq. (\ref{eq:m_r}) we write the stripped mass during each pericentric passage (Eq. \ref{eq:dm2}) as
\begin{equation}
\frac{\Delta M}{M_*} \simeq 4.8 f(M_*)^{3/2} \left[ 1 - \frac{f(M_0)}{f(M_*)} \frac{R_{\rm p}}{R_{\rm p0}} \left(\frac{M_*}{M_0}\right)^{2/3} \right]^{5/2}.
\label{eq:dm_evol}
\end{equation}
Here the dimensionless function $f(x) \equiv \left[1-(x/M_{\rm ch})^{4/3}\right]^{1/2}$ derives from Eq. (\ref{eq:m_r}), which is $\sim 1$ for $x \ll M_{\rm ch}$. Numerically solving Eqs. (\ref{eq:da_dt}-\ref{eq:dm_evol}), we can obtain the history of the mass stripping and the orbital evolution during the mass-loss stage. The results are shown in Figure \ref{fig:evolution}.

We can see from Figure \ref{fig:evolution} that, during the mass-loss stage, the orbital properties only have slight changes, i.e., $\dot e$ and $\dot R_{\rm p}$ are almost constant during the mass-loss stage. Consequently the fractional changes of the orbital properties are $\propto t$. Therefore, the duration of the mass-loss stage $t_{\rm ML}$, from the beginning of mass loss to the final disruption of WD, should be significantly shorter than $t_{\rm GW}$.

In the following, we analytically estimate $t_{\rm ML}$. The orbit-averaged mass-loss rate history can be written as $\dot M_* = -\Delta M / P$. Here we fix the orbital period to be the value at the beginning of the mass loss, i.e., $P \sim P_0$. The pericenter radius can be written as $R_{\rm p} \simeq R_{\rm p0} + \dot R_{\rm p0}t$, where $t=0$ is the onset of mass-loss and $\dot R_{\rm p0} = \dot R_{\rm p}|_{t=0} = -R_{\rm p0} /t_{\rm GW}$. Substituting this into Eq. (\ref{eq:dm_evol}) and taking the approximation $f(x) \simeq 1$, one obtains
\begin{equation}
\frac{\dot M_*}{M_0} \simeq - \frac{4.8}{P_0} \left\{1 - \left(1-\frac{t}{t_{\rm GW}}\right) \left[\frac{M_*(t)}{M_0} \right]^{2/3}\right\}^{5/2}.
\label{eq:dy_dt}
\end{equation}
Expanding the part in the curly brackets into a series of $(1-M_*/M_0)$, and taking its first order as long as $(M_0 - M_*)/M_0 \ll 1$, we obtain
\begin{equation}
\frac{\dot M_*(t)}{M_0} \simeq - \frac{4.8}{P_0} \left[\frac{t}{t_{\rm GW}} +  \frac{2}{3} \frac{(M_0 - M_*)}{M_0} \right]^{5/2},
\label{eq:dm_dt}
\end{equation}

Eq. (\ref{eq:dm_dt}) indicates two phases of the mass loss. At the early time, $M_*=M_0$, the mass loss is triggered and mainly driven by the orbital shrinkage due to GW radiation, thus the first term $t/t_{\rm GW}$ in Eq. (\ref{eq:dm_dt}) dominates. Later, the WD expands and its density drops after each mass loss, so the WD becomes easier to be tidally stripped. The mass loss will be very rapid, and soon the WD is fully disrupted. In this case, the second term $(2/3) (1-M_*/M_0)$ in Eq. (\ref{eq:dm_dt}) is $\gtrsim t/t_{\rm GW}$, and thus dominates. We define $t_{\rm tran}$ to be the transition time between these two phases. Thus, the mass-loss rate can be written as
\begin{equation}
\begin{split}
\frac{\dot M_*(t)}{M_0} &\sim -\frac{4.8}{P_0} \\
&\times \begin{cases}
\left(\frac{t}{t_{\rm GW}}\right)^{5/2},&\quad{t \lesssim t_{\rm tran}} \\
\left[\frac{2}{3} \left(1 - \frac{M_*}{M_0}\right) \right]^{5/2},&\quad{t \gtrsim t_{\rm tran}}.
\end{cases}
\end{split}
\label{eq:dm_dt2}
\end{equation}

We integrate Eq. (\ref{eq:dm_dt2}) from $t = 0$ to $t$ for $t\lesssim t_{\rm tran}$, and from $t$ to $t_{\rm ML}$ for $t\gtrsim t_{\rm tran}$, respectively, and use the boundary condition $M_*(t_{\rm ML}) = 0$. Then we obtain the cumulative mass loss fraction
\begin{equation}
\begin{split}
\frac{M_0 - M_*(t)}{M_0} \\
&\sim \begin{cases}
\frac{1.4}{P_0} t_{\rm GW}^{-5/2}t^{7/2},&\quad{t\lesssim t_{\rm tran}} \\
[1+2.6 \frac{(t_{\rm ML}-t)}{P_0}]^{-2/3},&\quad{t\gtrsim t_{\rm tran}},
\end{cases}
\end{split}
\label{eq:dm_GW_m}
\end{equation}
We can see that $\dot M_*/M_0 \propto t^{5/2}$ and $(M_0-M_*)/M_0 \propto t^{7/2}$ for $t \lesssim t_{\rm tran}$, which is consistent with the numerical result in Figure \ref{fig:evolution}. For most of the time when $t \gtrsim t_{\rm tran}$,  there should be $(t_{\rm ML} - t) / P_0 \gg 1$. Thus, the condition $(M_0 - M_*)/M_0 \ll 1$ is satisfied until the final disruption $t = t_{\rm ML}$, which can be seen in Figure \ref{fig:evolution}.

At $t = t_{\rm tran}$, the two effects that drive the mass loss, i.e., the orbit shrinkage and the $R_*$ expansion, are comparable. Thus, the two contributions to the mass-loss rate (Eq. \ref{eq:dm_dt2}) should be equal, i.e., $(2/3) [M_0-M_*(t_{\rm tran})]/M_0 = t_{\rm tran}/t_{\rm GW}$. The cumulative mass loss fraction (Eq. \ref{eq:dm_GW_m}) of the two phases should be equal too, i.e., $1 - M_*(t_{\rm tran})/M_0 = 1.4 t_{\rm tran}^{7/2} / (P_0 t_{\rm GW}^{5/2}) = [1+2.6 (t_{\rm ML}-t_{\rm tran})/P_0]^{-2/3}$. We then obtain
\begin{equation}
t_{\rm tran} \simeq P_0^{2/5} t_{\rm GW}^{3/5}.
\label{eq:t_tran}
\end{equation}
and
\begin{equation}
t_{\rm ML} \simeq 1.7 t_{\rm tran} - 0.4 P_0 \simeq 1.7 t_{\rm tran}.
\label{eq:t_ML0}
\end{equation}
The second equation is obtained by considering the fact that $t_{\rm tran} \gg P_0$. Thus, we have
\begin{equation}
\begin{split}
t_{\rm ML} &\simeq 0.002\ M_{\rm h,4}^{4/15} \left(\frac{M_0}{0.6\ M_{\odot}}\right)^{16/15} \ t_{\rm GW} \\
&\simeq 0.5 \, M_{\rm h,4}^{-2/5} \left(\frac{M_0}{0.6\ M_{\odot}}\right)^{-13/5} \left(\frac{1-e_0}{0.1}\right)^{-3/2} \ {\rm yr}.
\end{split}
\label{eq:t_ML}
\end{equation}
The duration of the mass-loss stage is much shorter than $t_{\rm GW}$ (cf. Eq. \ref{eq:t_GW}). This supports our earlier statement.


\subsection{Impact of the Mass Transfer}
\label{subsec:mass_transfer}
In the above where deriving the orbital evolution, we considered the impact of the GW radiation only and ignored the one due to the mass transfer. Considering the conservation of angular momentum, \cite{King_QPE_2022} assumes that the angular momentum is transferred to the WD while the mass is accreted onto the BH. As a result, the orbital pericenter radius should increase, as analogous to the orbital evolution of close binary systems \citep{Paczynski_Evolutionary_1971}.

Here we assess the effect of mass transfer on the orbital evolution, particularly for the case of the eccentric orbit in WD-IMBH system. Writing the time derivative of the orbital angular momentum $J = M_* \sqrt{GM_{\rm h} R_{\rm p} (1+e)}$ and rearranging it we obtain \citep[also see Eq. 10 of][]{King_QPE_2022}.
\begin{equation}
\frac{\dot R_{\rm p}}{R_{\rm p}} = -\frac{2\dot M_*}{M_*} + \frac{2\dot J}{J} - \frac{\dot e}{1+e}.
\label{eq:mass_transfer_expand}
\end{equation}

Two effects drive the orbital evolution in opposite directions: GW radiation causes the orbital shrinkage, with a rate given by Eq. (\ref{eq:t_GW}) or $(\dot R_{\rm p}/R_{\rm p})_{\rm GW} \simeq -t_{\rm GW}^{-1}$, while the mass transfer causes the orbit to expand.

To estimate the rate of the latter effect, we isolate it out and neglect the GW radiation for the moment. In addition, we consider the orbital angular momentum to be conserved, thus $\dot J = 0$. The mass transfer must be dissipative, i.e., it must be associated with some dissipation of the orbital energy, which tends to circularize the orbit, thus $\dot e < 0$. Therefore, the first and third terms on the r.h.s. of Eq. (\ref{eq:mass_transfer_expand}) are both positive. Furthermore, the magnitude of the third term should be comparable to or smaller than the first term -- otherwise, the term $2\dot M_*/M_*$ would be negligible, and then the circularization of the orbit would be dominated by other kind of effect than the mass transfer. Therefore, for the purpose of rate estimate, we can safely ignore the $\dot e$ term, and the orbital expansion rate via the effect of mass transfer is
\begin{equation}
\left(\frac{\dot R_{\rm p}}{R_{\rm p}} \right)_{\rm MT} \simeq -\frac{2\dot M_*}{M_*}.
\label{eq:dRp_mt}
\end{equation}
The real pericenter change rate would be the sum of the two:
\begin{equation}
\begin{split}
\frac{\dot R_{\rm p}}{R_{\rm p}} &= \left(\frac{\dot R_{\rm p}}{R_{\rm p}} \right)_{\rm MT} + \left(\frac{\dot R_{\rm p}}{R_{\rm p}} \right)_{\rm GW} \\
&\simeq -\frac{2\dot M_*}{M_*} - \frac{1}{t_{\rm GW}}.
\end{split}
\label{eq:mass_transfer_expand2}
\end{equation}

At the early time, the mass transfer rate is low, i.e., $-\dot M_*/M_* \ll t_{\rm GW}^{-1}$, such that the orbit expanding effect of mass transfer is unimportant and the orbit tends to shrink via GW radiation. In this case the mass loss rate increases monotonically ($\dot M_* \propto t^{5/2}$), as shown in Eq. (\ref{eq:dm_dt2}) and also in Figure \ref{fig:evolution}. However, at the late time, the mass transfer rate is high so its effect on the orbital evolution cannot be neglected. 

We can estimate the time when the effect of mass transfer becomes important. Substituting $\dot M_* / M_0 \simeq -t_{\rm GW}^{-1}$ into the first line of Eq. (\ref{eq:dm_dt2}), we have $4.8(t/t_{\rm GW})^{5/2}/P_0 \simeq t_{\rm GW}^{-1}$, and then we obtain $t \simeq 0.5\ P_0^{2/5} t_{\rm GW}^{3/5}$. Using Eq. (\ref{eq:t_tran}), we obtain that the effect of mass transfer is important when $t \gtrsim 0.5\ t_{\rm tran}$. This critical time and the critical mass tranfer rate ($t_{\rm GW}^{-1}$) are marked in Figure \ref{fig:evolution} as the two gray lines, respectively.

After that time, the mass loss would be slowed down, with respect to the results given in Eq. (\ref{eq:dm_dt2}-\ref{eq:dm_GW_m}) and Figure \ref{fig:evolution} where only the GW radiation is considered. The mass-loss stage would be prolonged, and the WD would be fully disrupted at a time much later than $t_{\rm ML}$ (Eq. {\ref{eq:t_ML}}). The detail of mass-loss evolution is crucial and deserves to be further explored in the future.

The above treatment is based on the conservation of orbital angular momentum, which derives from the classical circular-orbit close binary case, where the outer rim of the accretion disk around the accretor is continuously constrained by the the donor via the tidal interaction, and in the same interaction the angular momentum is relayed to the donor. However, the highly eccentric WD-IMBH system we consider here is different and more complicated. After an accretion disk around the BH forms with a circularization radius $\sim 2R_{\rm p}$ and viscously spreads outward, the angular momentum lost from the accreted material is transported to the outer part of the disk due to the viscosity. If the viscous timescale is shorter than the orbital period, the disk can freely spread outward while the WD is far from the pericenter. 

Another major difference is that in our case the WD is on a highly eccentric orbit may collide directly with the disk. When the WD returns to the pericenter and pass the disk, its velocity is faster than the Keplerian velocity of the disk gas at $R_{\rm p}$ by a factor of $\sqrt{2}$ \citep[see also in ][, their section 8.7]{Lu_QPE_2022}. A bow shock may be produced, where the ram pressure of the disk material in front of the WD reduces its orbital energy. In effect, it exerts a torque on the WD and decreases its angular momentum. After the passage, some of the shocked gas might escape the system, carrying away the angular momentum. 

Therefore, using Eq. (\ref{eq:dRp_mt}) -- which assumes that  the WD receives the angular momentum shed by the accretion -- may significantly overestimate the orbit expanding effect of the mass transfer in the highly eccentric WD-IMBH case. On the other hand, to our knowledge, those disk-WD interaction effects mentioned above have not been quantitatively explored for eccentric orbit cases, which merit a future investigation.

\section{detectability of EM and GW signals}
\label{sec:GW}

For the EM signal from a WD-IMBH system, if the peak luminosity is capped at the Eddington luminosity $L_{\rm Edd}$, the limiting distance for detection would be $d_{\rm lim, EM} \simeq [L_{\rm Edd}/(4\pi F_{\rm lim})]^{1/2} \simeq 323 (M_{\rm h}/10^4\ M_{\odot})^{1/2}$ Mpc. Here we adopted the limited flux for detection $F_{\rm lim} \simeq 10^{-13}\ {\rm erg\ s^{-1}\ cm^{-2}}$ \citep{Zhang_EPsensitivity_2022}, which is the sensitivity at $10$-s exposure time of the Follow-up X-ray Telescope (FXT) of the space X-ray mission Einstein Probe \citep[EP, ][]{Yuan_EP_2015}, to be launched at the end of 2023.

In the following we estimate the GW detectability. We begin by calculating the GW radiation from a WD inspiralling into IMBH in a eccentric orbit ($e \sim 0.7$ -- $0.9$) at the beginning of the mass-loss stage ($\beta \simeq 0.5$). The signal-to-noise ratio squared can be written as a summation of contributions from all harmonics of the orbital frequency \citep{Barack_Confusion_2004}, i.e., 
\begin{equation}
({\rm SNR})^2 = (\pi d)^{-2} \sum\limits_{n=1} \int \frac{4G\dot E_n}{c^3 f_n^2 S(f_n)} dt
\label{eq:SNR}
\end{equation}
where $d$, $f_n = n/P$ and $\sqrt{f_n S(f_n)}$ are the source distance, the GW radiation frequency and the sensitivity curve of GW detector, respectively, and
\begin{equation}
\dot E_n \simeq \frac{32}{5} \frac{G^4 M_*^2 M_{\rm h}^3}{c^5 a^5} g(n,e)
\label{eq:dotEn}
\end{equation}
is the power radiated in the $n$-th harmonics with $g(n,e)$ given by Eq. (20) in \cite{Peters_Gravitational_1963}.

\begin{figure*}		
\centering
\begin{minipage}[t]{0.5\textwidth}
\centering
 \includegraphics[scale=0.5]{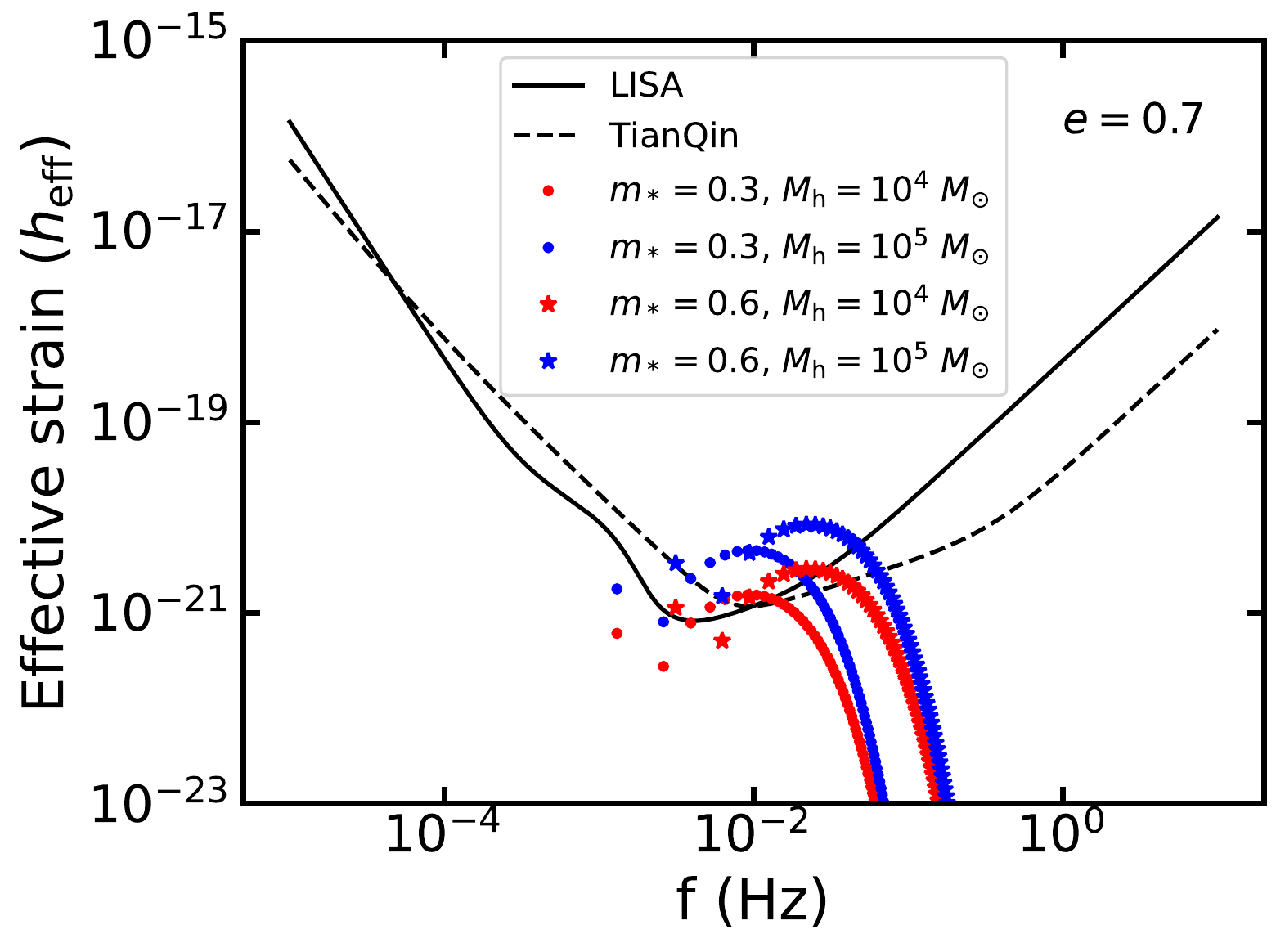}
 \end{minipage}
 \begin{minipage}[t]{0.48\textwidth}
\centering
 \includegraphics[scale=0.5]{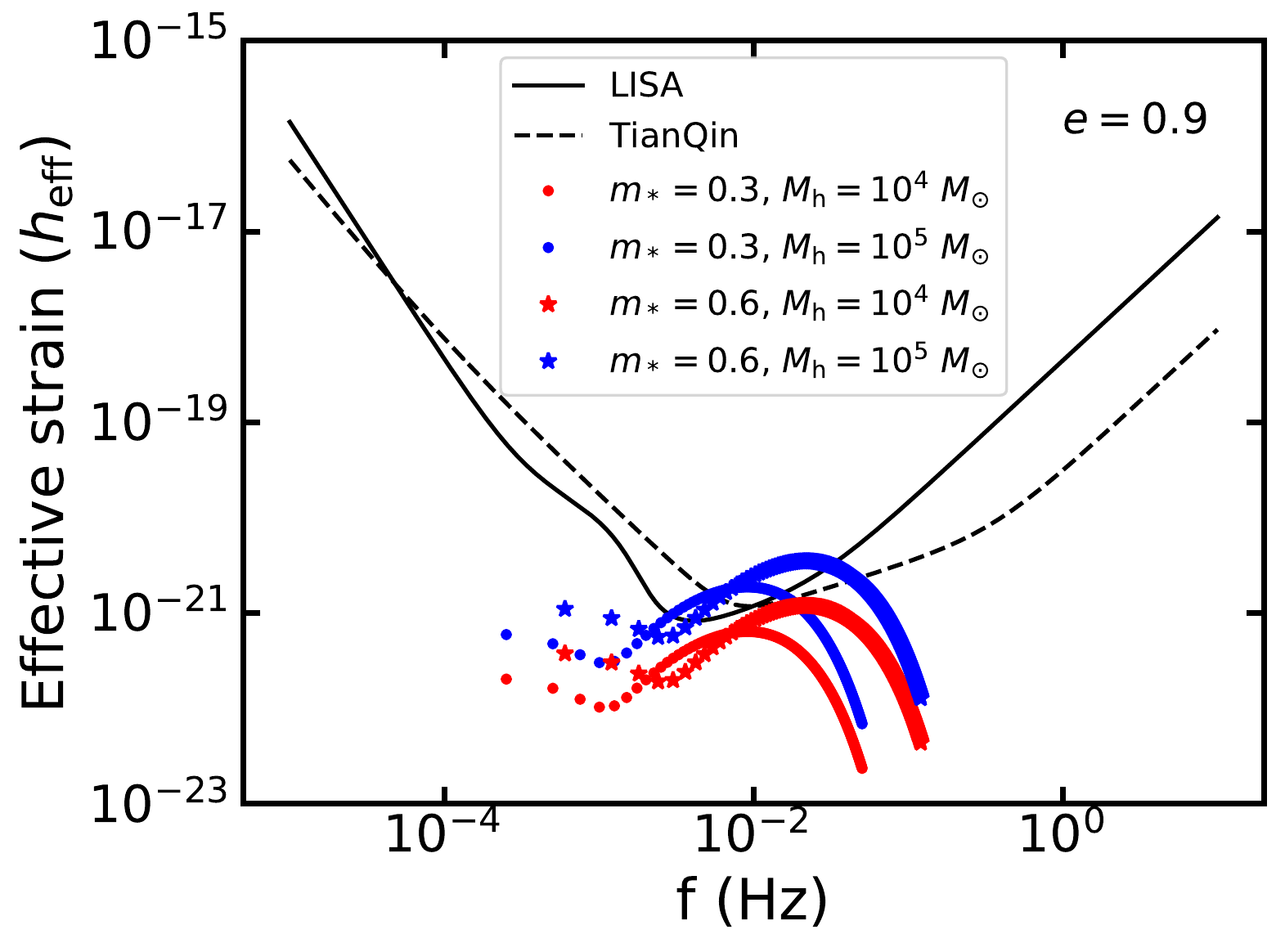}
 \end{minipage}
\caption{The effective GW strain (Eq. \ref{eq:heffn}) of WD-IMBH inspiral at the beginning of the mass-loss stage at a redshift of $z = 0.01$ ($d \simeq 43$ Mpc) versus the LISA \citep[solid line, ][]{Robson_The_2019} and TianQin \citep[dashed line, ][]{Luo_TianQin_2016} sensitivity curves. The colors and shapes represent different WD's masses and BH's masses. The orbital eccentricities are $e = 0.7$ and $e=0.9$ for the left and the right panels, respectively. The x axis for the effective GW strain denotes $f_n = n/P$ with $n=1,2,3...$.}
\label{fig:GW_SNR}
\end{figure*}

The timescale of the mass-loss stage $t_{\rm ML}$ is much shorter than $t_{\rm GW}$ (see Eq. \ref{eq:t_ML}). Therefore, the increment of frequency is negligible. One can assume the GW radiation frequency is equal to the value at the beginning of the mass-loss stage: $f_{\rm n} \simeq n/P_0$. Substituting Eq. (\ref{eq:dotEn}) into Eq. (\ref{eq:SNR}), we obtain
\begin{equation}
({\rm SNR})^2 \simeq \sum\limits_{n=1} \frac{h_{\rm eff,n}^2}{f_n S(f_n)} \simeq \frac{4G}{\pi^2 d ^2 c^3} \sum\limits_{n=1} \frac{\dot E_n T_{\rm obs}}{f_n^2 S(f_n)},
\label{eq:SNR2}
\end{equation}
where the observing duration $T_{\rm obs} \equiv \min({T_{\rm m}, t_{\rm ML}})$; $T_{\rm m} \simeq 4$ yr is the approximated mission lifetime of LISA and TianQin. The effective strain is defined as
\begin{equation}
h_{\rm eff,n} = \frac{2}{\pi d} \sqrt{\frac{G\dot E_n T_{\rm obs}}{c^3 f_n}}.
\label{eq:heffn}
\end{equation}
In Figure \ref{fig:GW_SNR}, we compare the effective strain for a redshift $z = 0.01$ ($d \simeq 43$ Mpc) WD-IMBH inspiral signal with the mission sensitivity curve $\sqrt{f_n S(f_n)}$ as an example.

For a WD-IMBH system with a highly eccentric orbit, its GW spectrum extents to a wide range of frequencies and peaks at the frequency \citep{Chen_MHz_2022}
\begin{equation}
\begin{split}
f_{\rm p} &= \frac{n_{\rm p}}{P_0} \simeq \frac{1}{2 \pi} \left(\frac{GM_{\rm h}}{R_{\rm p0}^3} \right)^{1/2} \\
&\simeq 2 \times 10^{-2}\ \left[\frac{f(M_0)}{f(0.6\ M_{\odot})} \right]^{-3/2} \left( \frac{M_0}{0.6\ M_{\odot}} \right)\ {\rm Hz},
\end{split}
\label{eq:f_p}
\end{equation}
with $n_{\rm p} \simeq (1-e)^{-3/2}$. Since the peak frequency $f_{\rm p}$ is of the same order as the Keplerian frequency of a circular orbit with the radius of $R_{\rm p0}$. It is independent from the BH's mass and falls in the sensitive band of LISA and TianQin ($\sim 10^{-2}$ Hz, see Figure \ref{fig:GW_SNR}).

Note that in this paper we are interested in those events that are simultaneously detectable in both the GW and EM windows. Actually the GW radiation emerges before the mass-loss stage during the inspiral. Therefore, these events are in priciple capable of cross verification.

In order to detect the combined GW and EM signals from the WD inspirals and the ensuing disruptions, the BH's and WD's masses should be in the intermediate range. If $M_{\rm h}$ or $M_*$ is too large, the WD will be directly swallowed by the BH rather than being disrupted; even when the mass loss stage occurs, its duration is too short to be detected. On the contrary, if $M_{\rm h}$ or $M_*$ is small, the GW signal will be too weak to be detected.

\begin{figure*}		
\centering
\begin{minipage}[t]{0.5\textwidth}
\centering
 \includegraphics[scale=0.5]{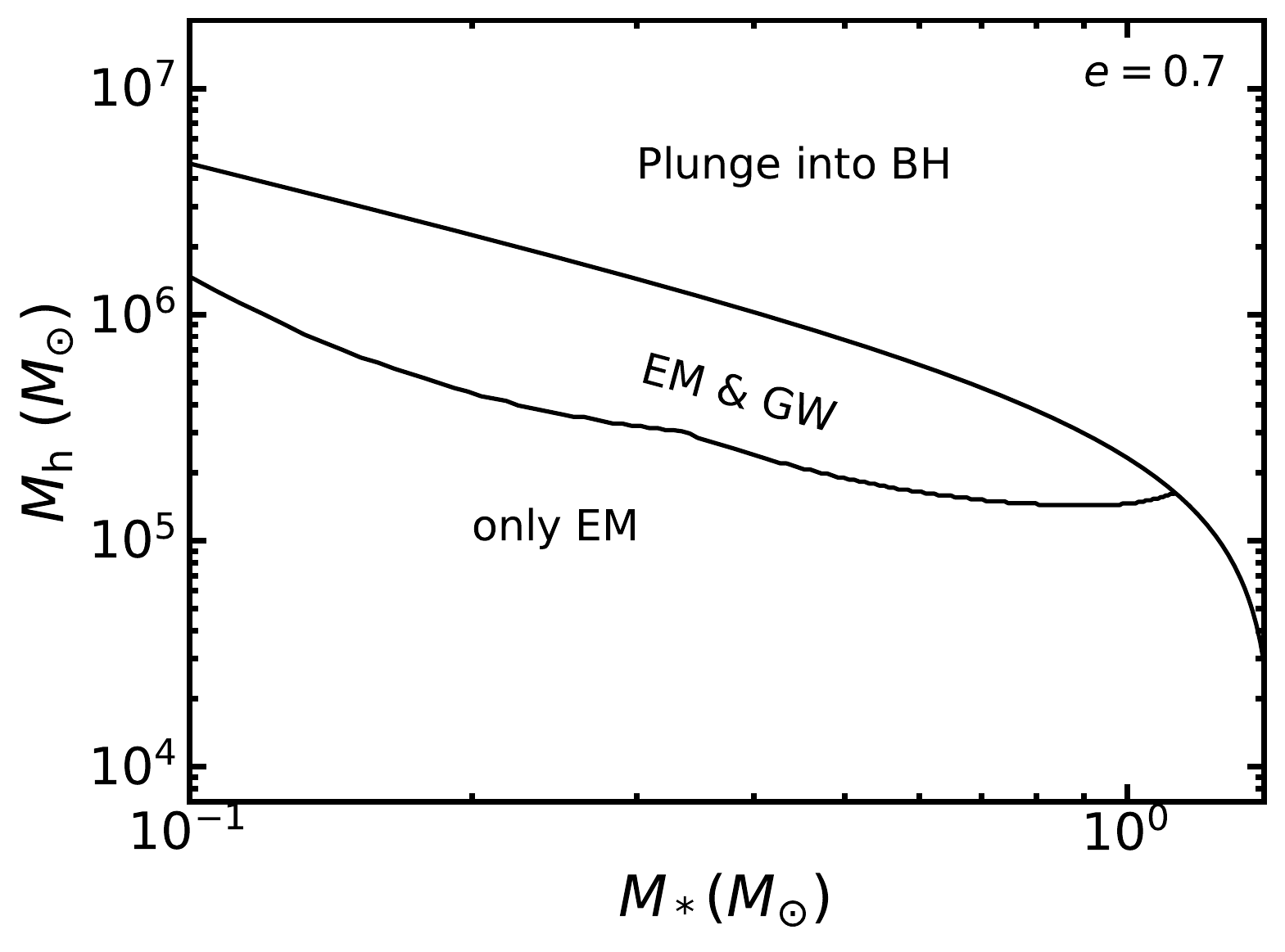}
 \end{minipage}
 \begin{minipage}[t]{0.48\textwidth}
\centering
 \includegraphics[scale=0.5]{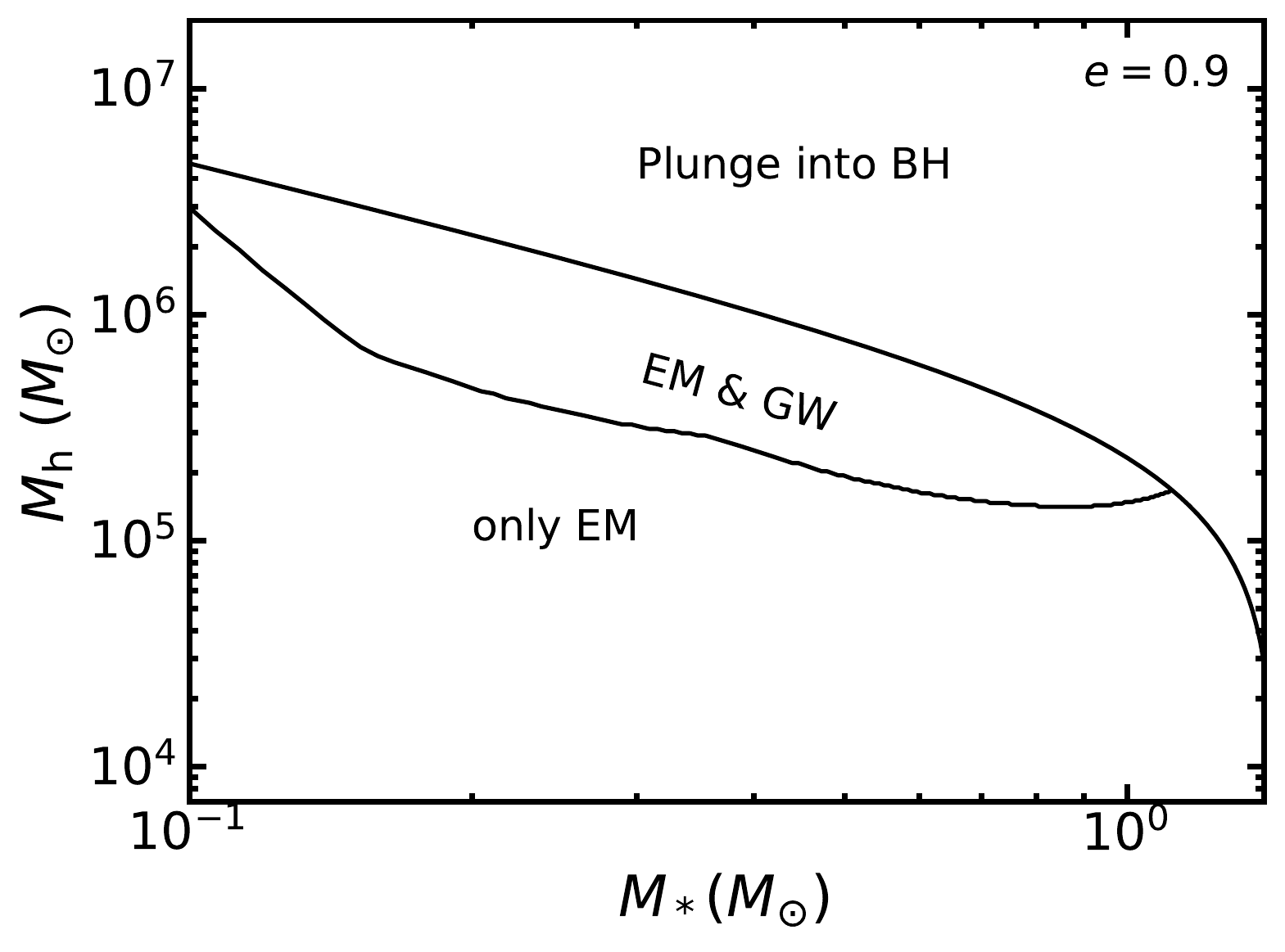}
 \end{minipage}
\caption{Detection prospect of the EM and GW signals in the parameter space of the BH mass and WD mass. The source redshift is taken to be $z=0.01$ ($d \simeq 43$ Mpc) and the eccentricity $e=0.7$ (left) and $e=0.9$ (right). We adopt the SNR threshold of 20 for GW detection. In the top region, $R_{\rm T}/\beta_0 \lesssim R_{\rm S}$, the WD would be directly plunge into the BH without mass loss during the inspiral. In the middle region, we can detect both of the EM and GW signals from the mass-loss stage. In the bottom region, we can only detect the EM signal.}
\label{fig:GW_diagram}
\end{figure*}

Figure \ref{fig:GW_diagram} illustrates the relevant parameter space for the detection of EM and GW signals, for a system  at a redshift of $z = 0.01$ ($d \simeq 43$ Mpc) with the eccentricity $e=0.7$ and $e=0.9$, respectively. Here and in the following, we adopt the SNR threshold of 20 for GW detection \citep{Babak_Science_2017,Fan_Science_2020}. We assume the WD will be directly swallowed by the BH and no mass loss during the inspiral occurs if $R_{\rm T}/\beta_0 \lesssim R_{\rm S}$. Figure \ref{fig:GW_diagram} implies that at this redshift, only those inspirals with BH mass $\sim 10^{5-6}\ M_{\odot}$ can be detected by both of the GW detector and the X-ray telescope.

\begin{figure*}		
\centering
\begin{minipage}[t]{0.5\textwidth}
\centering
 \includegraphics[scale=0.5]{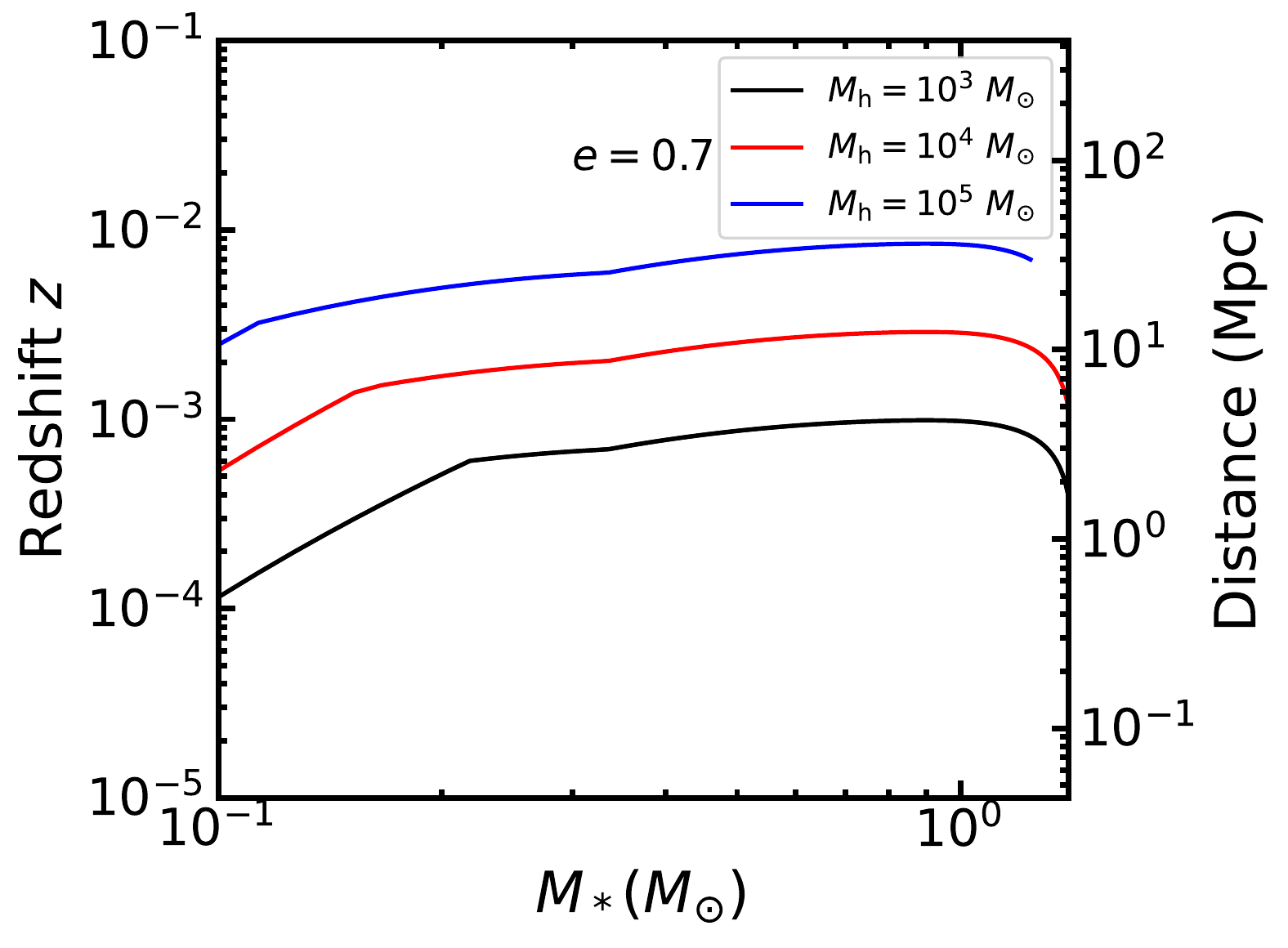}
 \end{minipage}
 \begin{minipage}[t]{0.48\textwidth}
\centering
 \includegraphics[scale=0.5]{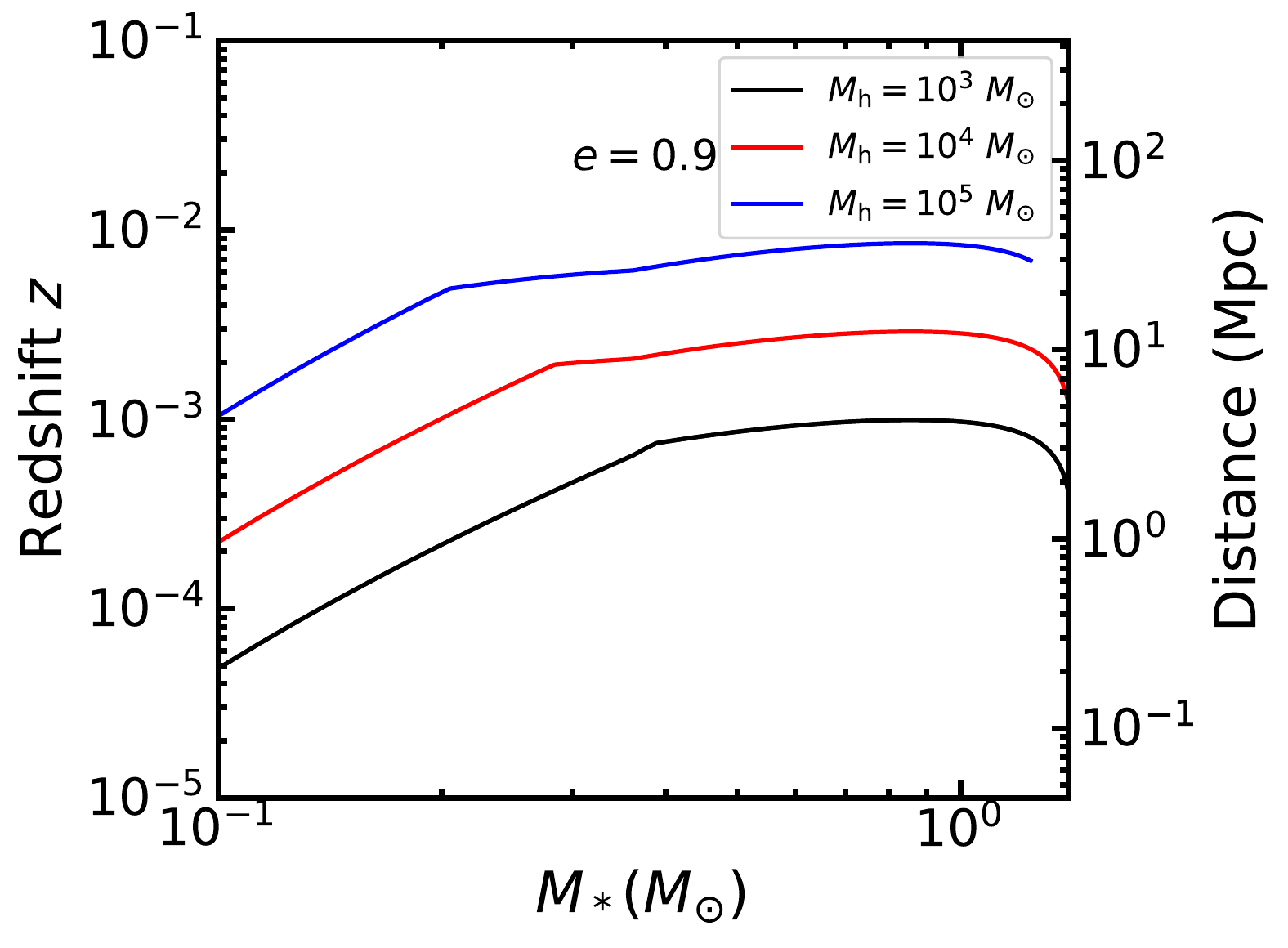}
 \end{minipage}
\caption{GW detectable horizon distance for the WD-IMBH inspiral during the mass-loss stage with the eccentricity $e=0.7$ (left) and $e=0.9$ (right). The colors represent different BH masses.}
\label{fig:GW_horizon}
\end{figure*}

The GW detectable horizon distance for WD-IMBH inspirals during the mass-loss stage is shown in Figure \ref{fig:GW_horizon}. Within the Local Supercluster ($\sim 33$ Mpc), we can detect those WD-IMBH inspiral with $M_{\rm h} \sim 10^5\ M_{\odot}$. For smaller IMBHs ($\sim 10^{3-4}\ M_{\odot}$), the detectable horizon distance is only $\sim 1 - 10$ Mpc. If these smaller IMBHs exist in star clusters of the nearby galaxies, we expect that they would be the ideal targets for the next-generation GW detectors. Furthermore, the detectable horizon distance peaks at $M_* \sim 1\ M_{\odot}$. This is because the duration of the mass-loss stage is too short for the WD-IMBH inspiral with massive WDs ($\gtrsim 1\ M_{\odot}$), and the GW signal is too weak for lighter WDs. 


\section{Conclusion}
\label{sec:conclusion}
WD-IMBH inspiral systems are simultaneously detectable in the GW and EM windows. Therefore, they provide a promising route for finding IMBHs, and a new probe for studying accretion physics and testing general relativity.

During the mass-loss stage of these systems, which are characterized by periodic EM bursts caused by accretion toward the IMBH, the total EM radiation energy budget depends on the accreted mass and in turn, on the total stripped mass $\Delta M$. If the eccentricity is small ($e \lesssim e_{\rm crit}$), all of the stripped mass would be bound to the IMBH and supply to the accretion disk. The duration of each burst is related to how and when the stripped mass joins into the disk. If the stripped mass supplies quickly to the disk as it returns to the pericenter, the properties of the fallback rate curve, e.g., $t_{\rm fb}$ and $t_{\rm peak}$, would be important for determining the light curve.  

We analytically calculate the parameter dependence of the stripped mass $\Delta M$ of the WD, the critical eccentricity $e_{\rm crit}$, and the mass fallback timescale $t_{\rm fb}$. We then perform hydrodynamic simulations to verify them. The results are as follows:
\begin{itemize}
\item The stripped mass $\Delta M$ depends only on the impact factor $\beta$ and the WD mass $M_*$ (Eq. \ref{eq:dm2}). For larger $M_*$, because of a higher central mass concentration in the WD, the fractional of stripped mass $\Delta M/ M_*$ is smaller. The simulation results are consistent with the analytical one for $\beta \lesssim 0.7$. However, we find from the simulation that the orbital eccentricity $e$ slightly affects the stripped mass: $\Delta M$ is slightly larger for smaller $e$.

\item We derived a critical eccentricity $e_{\rm crit}$, at which all of the debris are bound to the IMBH. It depends on the mass ratio $M_{\rm h} / M_*$ and $\beta$ and also on the parameter $\xi$ that quantifies the tidal spin-up effect on the WD upon the pericenter passage (see Eq. \ref{eq:E_tail2}). The simulation shows that $e_{\rm crit}$ is smaller for the larger WD, consistent with the analytical calculation. Using the simulations results, we can constrain $\xi$ to be $\sim 3.5$ -- $4$.

\item The fallback time $t_{\rm fb}$ for the most bound of the stripped material is a fraction of the WD orbital period $P$ (Eq. \ref{eq:t_fb} and Figure \ref{fig:ratio_tfb}). Since the stripped mass joins the accretion disk only when it returns to the pericenter, the outburst may appear only when $t \gtrsim t_{\rm fb}$ after each pericentric encounter. The simulation shows that the fraction $t_{\rm fb} / P$ is smaller for larger $M_*$ and higher $e$, consistent with the analytical formula Eq. (\ref{eq:t_fb}).

\item From the hydrodynamic simulation result we infer the mass fallback rate of the stripped mass. Because the debris stream is truncated by the survived WD, the mass fallback rate has a dip when the WD returns to the pericenter. We find that the first peak of the fallback rate is at least one order of magnitude higher and more remarkable than the later one. If the debris can be rapidly accreted by the IMBH when it returns to pericenter, the light curve should evolve mimicking the mass fallback rate.

\item From the simulation results, we obtain an empirical fitting formula for the rise time $t_{\rm rise}$ of the fallback rate, i.e., Eq. (\ref{eq:Gamma2_fit}), which shows that $t_{\rm rise} / t_{\rm fb}$ is larger for larger $M_*$ and $e$.

\end{itemize}

Using the estimate of the stripped mass (Eq. \ref{eq:dm2}), we calculate the long-term evolution of the mass-loss stage of the system. The mass loss is initially driven by the orbital shrinkage due to GW radiation, and the orbital mean of the mass loss rate grows as $\dot M_* \propto t^{5/2}$. Later when the WD expands, the mass loss becomes drastic and the WD is fully disrupted at $t \simeq t_{\rm ML}$. The duration of the mass-loss stage $t_{\rm ML}$ is typically a few months to years, a small fraction of the WD orbit shrinkage timescale $t_{\rm GW}$ that is purely due to the GW radiation.

We also consider the orbital expansion due to the mass transfer, under the assumption of orbital angular momentum conservation. We find that this effect would make the mass loss stage prolonged, hence the WD would be fully disrupted at a time much later than $t_{\rm ML}$ (where this effect was not considered).

We calculate the detectability of EM and GW signals during the mass-loss stage of the WD-IMBH inspiral. The EM signal can be easily detected if the peak luminosity of each burst is Eddington luminosity. The GW signal is detectable only within a distance of $1 - 50$ Mpc, depending on the IMBH and WD masses. Those WD-IMBH inspirals with larger WD masses ($\sim 1\ M_{\odot}$) will be the ideal targets for the GW detection by LISA and TianQin. 

Notice that most of the observed QPEs have been considered to be the tidal stripping with high eccentricity $e > 0.9$, or even as extreme as $e \simeq 0.99$ \cite[e.g., ][]{Chen_MHz_2022}. However, such highly eccentric tidal stripping is difficult to simulate, because its orbital period is very long and it is too time-consuming. Hence, the most eccentric case that we simulated is $e=0.9$. Those extremely eccentric cases that are particularly relevant to QPEs remain a subject of future simulations.


\section{acknowledgments}
J.H.C thanks Dr. Guobin Mou for the devoted instruction of hydrodynamic simulation. This work is supported by the National Natural Science Foundation of China under grant Nos. 12073091 and 11903089, the Guangdong Basic and Applied Basic Research Foundation under grant Nos. 2019A1515011119, 2021B1515020090 and 2019B030302001, the Fundamental Research Funds for the Central Universities, Sun Yat-sen University under grant No. 22lgqb33, and the China Manned Space Project under grant Nos. CMS-CSST-2021-A11 and CMS-CSST-2021-B09. The FLASH code used for the simulations in this work was developed in part by the DOE NNSA, and DOE Office of Science supported Flash Center for Computational Science at the University of Chicago and the University of Rochester.

\clearpage

\begin{appendix}
\label{Appendix}
\section{Structure of white dwarfs and the modified Lane-Emden equation}
\label{subsec:LE}
In our paper, we consider a wide range of the WD mass. Although the density profile of a small ($\simeq 0.2\ M_{\odot}$) or a massive ($\gtrsim 1\ M_{\odot}$) WD can be approximated by a spherical polytrope with $\gamma = 5/3$ and $\gamma = 4/3$, respectively, strictly speaking, most of the WDs' structure are different from these two polytropes. Therefore, we need to calculate a more general density profile of a WD, which is governed by the elctron-degenerate equation of state. We then use it in our simulation initial setup. In the following, we recalculate the WD's structure following \cite{Chandrasekhar_Stellar_1935}.

The general expression for electron-degenerate pressure is
\begin{equation}
P \simeq P_0 \int_0^{x_{\rm F}} \frac{x^4}{(1+x^2)^{1/2}}\ dx, 
\label{eq:P}
\end{equation}
where 
\begin{equation}
P_0 \equiv \frac{8\pi m_{\rm e}^4 c^5}{3 h^3},
\label{eq:P0}
\end{equation}
and $m_{\rm e}$, $h$ and $c$ are the electron mass, Planck constant and the light speed, respectively; $x \equiv p / (m_{\rm e} c)$ is the dimensionless momentum. The Fermi momentum is given by
\begin{equation}
x_{\rm F} = \left( \rho/\rho_0\right)^{1/3},
\label{eq:rho}
\end{equation}
where $\rho$ is the gas density and
\begin{equation}
\rho_0 \equiv \frac{8\pi}{3} \mu_{\rm e} m_{\rm p} \left (\frac{m_{\rm e} c}{h} \right)^3 \simeq 2 \times 10^6 \ {\rm g\ cm^{-3}}
\label{eq:rho0}
\end{equation}
is the critical density that separates the relativistic ($x_{\rm F} \gtrsim 1$) case from the non-relativistic ($x_{\rm F} \ll 1$) case for the electrons. The mean molecular weight per electron is $\mu_{\rm e} \simeq 2$ for typical WD's composition.

\cite{Chandrasekhar_Stellar_1935} has thoroughly studied the equation of state of a degenerate star, and he derived a modified Lane-Emden equation to describe a WD. Here, we repeat this derivation.

The WD is in hydrostatic balance, thus we have
\begin{equation}
\frac{dP}{dr} = - \rho \frac{G M(r)}{r^2}.
\label{eq:dP_dr}
\end{equation}
Here $M(r) = \int_0^r 4 \pi r^2 \rho\ dr$ is the mass contained within radius $r$. The chain rule gives
\begin{equation}
\frac{dP}{\rho dr} = \frac{1}{\rho} \frac{dP}{dx_{\rm F}} \frac{dx_{\rm F}}{dr}.
\label{eq:dP_dxf}
\end{equation}
Substituting Eq. (\ref{eq:P}) and Eq. (\ref{eq:rho}) into Eq. (\ref{eq:dP_dxf}), we obtain
\begin{equation}
\begin{split}
\frac{dP}{\rho dr} &= \frac{P_0}{\rho_0}\frac{d(1+x_{\rm F}^2)^{1/2}}{dr} \\
&= -G\rho_0 \frac{\int^r_0 4 \pi r'^2 x_{\rm F}(r')^3\ dr'}{r^2}.
\end{split}
\label{eq:dP_dr2}
\end{equation}

Taking the derivative of Eq. (\ref{eq:dP_dr2}) with respect to $r$, we obtain $(P_0/\rho_0) d^2(1+x_{\rm F}^2)^{1/2}/dr^2 = -G \rho_0 [4\pi x_{\rm F}^3 - (2/r^3) \int_0^r 4\pi r'^2 x_{\rm F}(r')^3\ dr'] = -4 \pi G \rho_0 x_{\rm F}^3 - (2P_0)/(r \rho_0) d(1+x_{\rm F}^2)^{1/2}/dr$. Defining $\phi \equiv (1+x_F^2)^{1/2} / (1+x_0^2)^{1/2}$ and a dimensionless length $\zeta \equiv r/\alpha$, one obtains the modified Lane-Emden equation
\begin{equation}
\frac{1}{\zeta^2} \frac{d}{d\zeta}\left( \zeta^2 \frac{d\phi}{d\zeta} \right) = - \left( \phi^2-\frac{1}{1+x_0^2}\right)^{3/2}.
\label{eq:L-E}
\end{equation}
Here $x_0$ is the central value of $x_{\rm F}$, i.e., $x_0 = x_{\rm F}(\zeta=0)$, and 
\begin{equation}
\alpha = \left[\frac{P_0}{4\pi G \rho_0^2 (1+x_0^2)} \right]^{1/2}.
\label{eq:alpha}
\end{equation}

\begin{figure}		
\centering
 \includegraphics[scale=0.5]{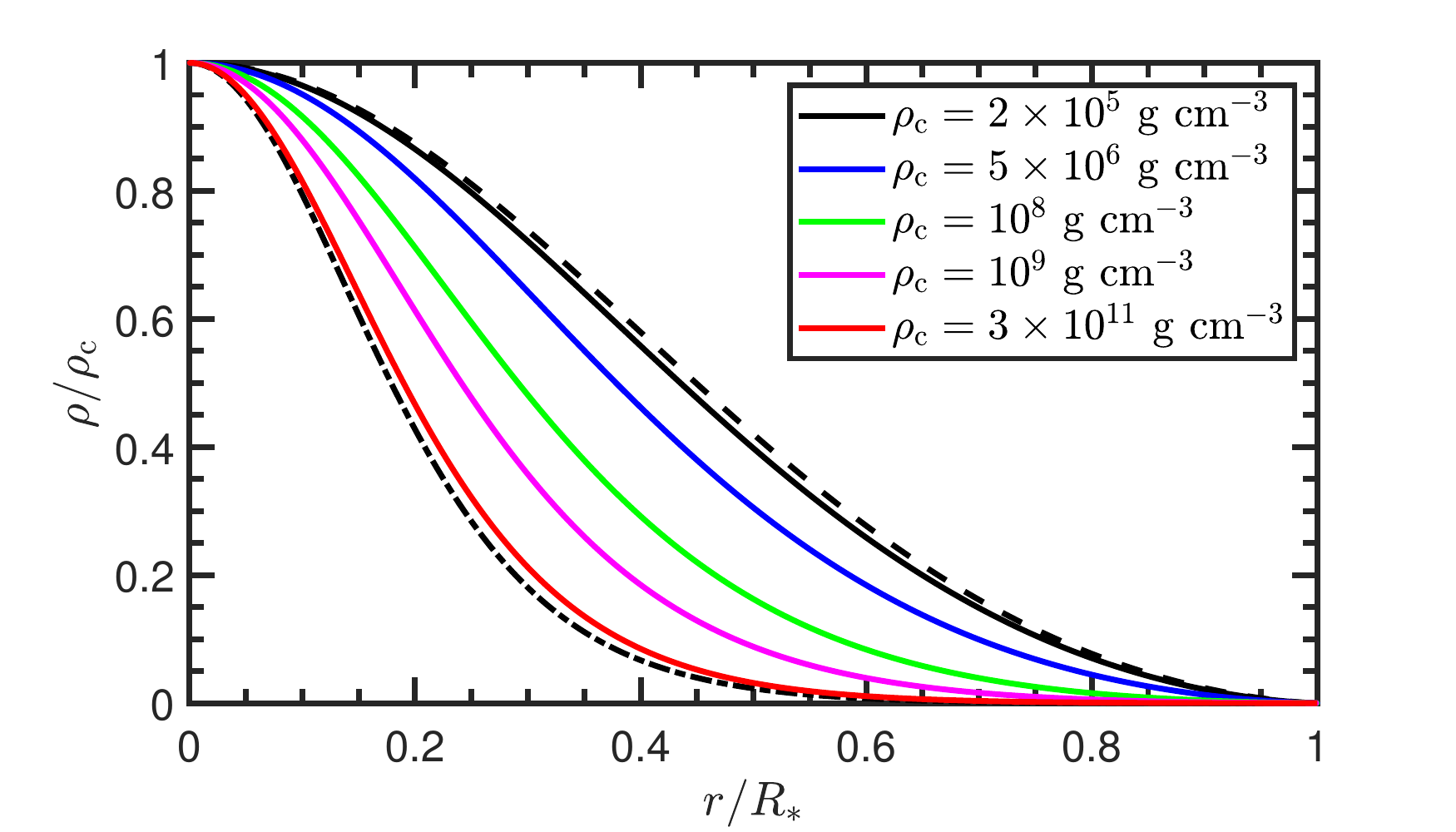}
\caption{Density profiles of the WDs with different central density. The black dashed and dot-dashed lines are the density profiles of $\gamma = 5/3$, $4/3$ spherical polytropes, respectively. The least and most massive WDs can be approximated by the polytropes with $\gamma = 5/3$ and $\gamma = 4/3$, respectively. However, the density profile of most of the WDs cannot depicted by a single polytrope.}
\label{fig:profile}
\end{figure}

We input several central densities into Eq. (\ref{eq:L-E}), ranging from $2 \times 10^5\ {\rm g\ cm^{-3}}$ to $3 \times 10^{11}\ {\rm g\ cm^{-3}}$, to calculate the WD's density profiles and plot them in Figure \ref{fig:profile}. The lowest central density is chosen to be the approximate value for the lowest-mass WD ($\lesssim 0.2\ M_{\odot}$) that have been found \citep{Kilic_LowestWD_2007}. The highest central density is about the onset of converting protons to neutrons (the so-called neutron drip).

More massive WDs have higher densities. For a WD with high density, electrons in its most part are relativistic ($x_{\rm F} \gtrsim 1$), thus, the degenerate pressure (Eq. \ref{eq:P}) is $P \propto x_{\rm F}^4 \propto \rho^{4/3}$. On the contrary, WDs with lower densities are sub-relativistic, so $P \propto x_{\rm F}^5 \propto \rho^{5/3}$. Figure \ref{fig:profile} shows that the least and the most massive WDs can be approximated by a $\gamma = 5/3$ and a $\gamma = 4/3$ polytropes, respectively. However, for most of the WDs with a mass within that range, a single polytrope cannot depict the density profile. That is because the electron gas of the WD is relativistic inside and non-relativistic on the surface. Therefore, it is necessary to consider a more accurate profile, as given by Eq. (\ref{eq:L-E}), in hydrodynamic simulations. It is more so when we need to determine the tidally stripped mass of a WD, which is sensitive to the WD's structure \citep{Liu2013}.

\end{appendix}

\clearpage
\bibliography{cited}

\end{CJK*}
\end{document}